\documentclass[10pt,journal]{IEEEtran}
\usepackage[left=1in,right=1in,top=1in,bottom=1in]{geometry}
\usepackage{tikz}
\usetikzlibrary{arrows.meta}
\usepackage{amsmath,amssymb,amsfonts}
\usepackage{amsthm}
\usepackage{bm}
\usepackage{mathtools}
\usepackage{cite}
\usepackage{graphicx}
\usepackage[hidelinks]{hyperref}
\usepackage{orcidlink}
\hypersetup{pdftitle={Finite-State Uncertainty-Aware Widely-Linear Equalization for Aeronautical GMSK Links},pdfauthor={Ahmad Tanha and Ali Khalesi}}

\newtheorem{theorem}{Theorem}
\newtheorem{corollary}{Corollary}
\newtheorem{proposition}{Proposition}

\newcommand{\E}{\mathbb E}
\newcommand{\CN}{\mathcal{CN}}
\newcommand{\tr}{\operatorname{tr}}
\newcommand{\bfeps}{\bm\epsilon}
\newcommand{\bfth}{\bm\theta}

\newtheorem{lemma}{Lemma}

\begin{document}

\bstctlcite{IEEEexample:BSTcontrol}


\title{Finite-State Uncertainty-Aware Widely-Linear Equalization for Aeronautical GMSK Links}

\author{Ahmad Tanha~\orcidlink{0000-0003-3227-8667},~\IEEEmembership{Member,~IEEE} and Ali Khalesi~\orcidlink{0000-0002-5815-3611},~\IEEEmembership{Member,~IEEE}
\thanks{
The authors are with the Signals, Telecommunications \& AI Department at Institut Polytechnique des Sciences Avancées (IPSA), Ivry-sur-Seine, France. Ali Khalesi is also with LINCS lab, Paris, France.
Emails: \{ahmad.tanha, ali.khalesi\}@ipsa.fr.
}
}

\maketitle
\begin{abstract}
Aeronautical and unmanned aerial vehicle (UAV) receivers must
equalize doubly selective channels under synchronization
uncertainty and limited onboard computation and memory.
We propose a finite-state, uncertainty-aware widely-linear
(WL) equalization algorithm for Gaussian minimum-shift keying
(GMSK) links. The algorithm uses synchronization estimates
and their reliability descriptors to select one of a small
number of precomputed filters, avoiding repeated online
coefficient synthesis. Using a finite-memory Laurent model,
we design the stored affine WL filters from conditional
physical-state moments. We derive an exact characterization
of excess mean-square error (MSE) relative to adaptation
to continuous side information, Lloyd-type conditions for
finite-state design, local timing--frequency mismatch
certificates, and an orthogonal decomposition of design
and information losses. On a reference channel at
$E_b/N_0=8$~dB under high synchronization-estimation
uncertainty, a $16$-state bank reduces pseudo-symbol MSE
by $51.9\%$ relative to point-estimate WL time-varying
MMSE equalization and by $6.6\%$ relative to posterior-mean
(PM) substitution. With identical partitioning, routing,
and storage, conditional-moment filters reduce MSE
by $10.1\%$ and full-waveform average bit error rate
(ABER) by $15.6\%$ relative to representative-state
filters. Across $18$ synthetic three-path channels with
individually fitted banks, MSE and ABER decrease by
$5.3\%$ and $12.9\%$, respectively, relative to PM
substitution. The $16$-state receiver stores $2048$ real
coefficients. Posterior integration, robustness sweeps,
and software timing measurements quantify the tradeoff
between equalization accuracy, online computational cost,
and bank refresh.
\end{abstract}

\begin{IEEEkeywords}
Aeronautical communications, GMSK, widely-linear equalization, synchronization uncertainty, doubly selective channels, MMSE estimation, finite-state receivers.
\end{IEEEkeywords}

\section{Introduction}
\label{sec:introduction}

Aeronautical communication links, including unmanned aerial
vehicle (UAV) control and non-payload links, require robust
receivers with limited onboard computation and memory~\cite{Darsena2018}.
Relative motion and multipath produce doubly selective channels,
while imperfect front-end estimation leaves residual timing
and frequency uncertainty. Repeated equalizer coefficient
synthesis can impose appreciable online cost. We therefore
focus on sampled linear and widely-linear (WL) equalizers
and propose a finite-state implementation that uses
synchronization estimates and reliability information
to select uncertainty-aware filters from a precomputed bank.

Gaussian minimum-shift keying (GMSK) has a constant
envelope and admits a Laurent continuous-phase
modulation (CPM) representation~\cite{Murota1981,Laurent1986}.
The resulting pseudo-symbols can be improper, motivating
WL processing~\cite{Picinbono1995,Gerstacker2003,Darsena2017SOS}.
Conjugate observations allow WL processing to exploit
these statistics within an affine filtering architecture.
Recent work includes WL minimum mean-square error (MMSE)
receivers for rectilinear and quasi-rectilinear signals~\cite{chevalier2024new},
low-complexity WL equalizers for spectrally shaped discrete
Fourier transform-spread orthogonal frequency-division
multiplexing~\cite{chae2023low}, and satellite-impairment
compensation~\cite{luo2025widely}. We focus on uncertainty-aware
adaptation under a finite storage budget.

The closest application-specific work develops linear
time-varying (LTV) and widely-linear time-varying (WLTV)
zero-forcing and MMSE equalizers for CPM over doubly
selective UAV links~\cite{Darsena2018}.
Our point-estimate plug-in (PI) WLTV-MMSE baseline treats
the estimated physical state as exact. We instead design stored
filters from covariance and complementary-covariance moments
conditioned on the estimates and reliability descriptors
assigned to each receiver state. This targets conditions in
which synchronization uncertainty degrades point-estimate
adaptation but side information still supports effective
filter selection. The numerical evaluation includes path
Doppler and residual timing--frequency mismatch.

Low-complexity CPM equalization has also exploited
frequency-domain band structure~\cite{Chayot2018Band}.
Related approaches partition decoder side information
using context trees with local linear turbo equalizers
and MMSE-related guarantees~\cite{Kim2013Context},
or cluster reliability information with local adaptive
equalizers~\cite{Kim2014Clustering}.
Our partition captures physical-state uncertainty, including
synchronization and channel parameters. A matched-budget
experiment isolates conditional-moment filter-design gains
with partition, routing, and storage fixed.

Bayesian equalization can account for channel
uncertainty~\cite{Salamanca2012Bayesian}, and complex
Gaussian-process regression relates Bayesian inference
to WL MMSE estimation~\cite{boloix2024complex}.
Unlike observation quantization~\cite{Shlezinger2019Task},
our approach leaves the sampled observation unchanged
and compresses receiver adaptation.
Specifically, we distinguish the physical state
$\bfth=(\bfeps,\mathbf h,\mathbf R_n^{\rm true})$
from the available side information
$\mathsf S=(\widehat{\bfeps},\widehat{\mathbf h},
\widehat{\mathbf R}_n,\boldsymbol\zeta)$
and use $p(\bfth\mid\mathsf S)$ to compute conditional
moments.
Building on classical affine-MMSE projection and
Lloyd/Linde--Buzo--Gray (LBG) alternation, we develop
a finite-state adaptation framework for synchronization
uncertainty.

Our main contributions are summarized below.
\begin{itemize}
\setlength{\itemsep}{1pt}
\setlength{\parskip}{0pt}
\setlength{\parsep}{0pt}

\item We formulate a finite-state affine-WL equalizer
under a fixed storage budget.
State estimates and reliability descriptors select
precomputed filters designed from conditional
physical-state moments, avoiding repeated online
coefficient synthesis.

\item We characterize the exact finite-state MSE
penalty relative to adaptation to continuous side
information and specialize classical Lloyd/LBG
conditions to excess equalization MSE.
We distinguish the threshold-based bank evaluated
numerically from unrestricted Lloyd optimization.

\item We characterize the side-information and
oracle affine-WL limits, decompose plug-in degradation
into orthogonal design and information losses,
and derive local timing--frequency viability
certificates for frozen WLTV equalization relative
to the corresponding complex-linear receiver.

\item We demonstrate an accuracy--computation tradeoff
for resource-limited airborne receivers in terms of MSE and
full-waveform average bit error rate (ABER). Reference-channel,
multi-channel, and matched-budget experiments assess the
benefit of conditional-moment filter design.
Comparisons with continuous adaptation, robustness tests,
and software timings quantify accuracy, online cost,
and bank-refresh requirements.

\end{itemize}

\textbf{Notation.}
Bold lowercase and uppercase letters denote vectors
and matrices, respectively; $(\cdot)^T$, $(\cdot)^*$,
and $(\cdot)^H$ denote transpose, complex conjugation,
and Hermitian transpose.
The operators $\E[\cdot]$, $\operatorname{Cov}(\cdot)$,
$\tr(\cdot)$, and $\|\cdot\|_2$ denote expectation,
covariance, trace, and Euclidean norm, respectively.
For Hermitian matrices, $\mathbf A\succeq\mathbf B$
means that $\mathbf A-\mathbf B$ is positive
semidefinite, and $\CN(\boldsymbol\mu,\mathbf R)$
denotes a proper complex Gaussian distribution
with mean $\boldsymbol\mu$ and covariance $\mathbf R$.
Underlined variables denote augmented WL vectors;
$\mathsf S$ and $\mathsf O$ denote side information
and oracle information, respectively.

\section{System Model and Problem Formulation}
\label{sec:system_model}

We model a doubly selective GMSK link with imperfect
front-end information and formulate uncertainty-aware
adaptation within the affine widely-linear (WL)
receiver class.

\subsection{GMSK Block Representation}
\label{subsec:gmsk_model}

Let $\{a_m\}$, $a_m\in\{-1,+1\}$, denote a binary
GMSK sequence with symbol duration $T$.
GMSK is a binary CPM with modulation index $h=1/2$
and Gaussian frequency pulse shaping~\cite{Murota1981}.
Using Laurent's decomposition~\cite{Laurent1986},
the adopted finite-memory GMSK waveform over the
observation interval can be represented as
\begin{equation}
s(t)=\sum_{k=1}^{K} b_k g_k(t),
\label{eq:compact_laurent}
\end{equation}
where $k$ indexes a Laurent component and its time shift,
$g_k$ is the corresponding pulse, and $b_k$ is its
data- and phase-state-dependent pseudo-symbol.
The analysis uses the exact representation of the
adopted finite-memory model over the observation
interval; dominant-component truncation is used
only in the numerical implementation.

We collect the pseudo-symbols as
\begin{equation}
\mathbf b
=
\begin{bmatrix}
b_1 & \cdots & b_K
\end{bmatrix}^{T}.
\label{eq:b_vector}
\end{equation}
We allow $\mathbf b$ to have a nonzero mean.
Let $\boldsymbol\mu_b=\E[\mathbf b]$, and define
its centered covariance and pseudo-covariance as
\begin{align}
\mathbf R_b
&=
\E\!\left[
(\mathbf b-\boldsymbol\mu_b)
(\mathbf b-\boldsymbol\mu_b)^H
\right],\\
\mathbf C_b
&=
\E\!\left[
(\mathbf b-\boldsymbol\mu_b)
(\mathbf b-\boldsymbol\mu_b)^T
\right].
\label{eq:b_statistics}
\end{align}
Depending on the pseudo-symbol statistics and phase
initialization, $\mathbf C_b$ may be nonzero,
motivating WL processing.

\subsection{Aeronautical Channel and Residual Synchronization}
\label{subsec:physical_channel}

The channel has $L_h$ paths with delays $\lambda_\ell$
and time-varying coefficients
$\alpha_\ell[n]$~\cite{Bello1963,Darsena2018}.
For sampling period $T_s$ and known phase-reference
start time $t_0$, sample $n$ has local time $nT_s$
and phase-reference time $t_0+nT_s$,
$n=0,\ldots,N-1$.
We write
\begin{equation}
\alpha_\ell[n]
=
\bar{\alpha}_\ell[n]
e^{j2\pi\nu_\ell(t_0+nT_s)},
\label{eq:path_doppler}
\end{equation}
where $\nu_\ell$ is the path Doppler shift and
$\bar\alpha_\ell[n]$ captures the remaining gain variation.
Imperfect synchronization leaves a residual timing
error $\Delta\tau$ and a common frequency offset
$\Delta f$.
Unlike path-dependent Doppler, $\Delta f$ produces
the same phase rotation on all paths.

We normalize the residual synchronization errors as
\begin{equation}
\epsilon_\tau=\frac{\Delta\tau}{T},
\qquad
\epsilon_\nu=\Delta f\,T,
\qquad
\bfeps=
\begin{bmatrix}
\epsilon_\tau\\
\epsilon_\nu
\end{bmatrix}.
\label{eq:sync_state}
\end{equation}

For receive filter $r(t)$, let $\phi_k=g_k*r$
denote the nominal effective Laurent pulse.
We adopt a sampled-equivalent phase model in which
path Doppler and residual frequency offset act as
phase rotations at the sampling instants.
Channel phases use $t_0+nT_s$, while pulse arguments
use local time.
The resulting $N\times K$ channel matrix is
\begin{equation}
\begin{split}
[\mathbf H(\bfeps,\mathbf h)]_{n,k}
&=
\sum_{\ell=0}^{L_h-1}
\bar{\alpha}_\ell[n]\,
e^{j2\pi(t_0+nT_s)
\left(\nu_\ell+\epsilon_\nu/T\right)}
\\[-1mm]
&\times
\phi_k\left(nT_s-\lambda_\ell-T\epsilon_\tau\right),
\end{split}
\label{eq:true_channel_matrix}
\end{equation}
where
\begin{equation}
\mathbf h
\triangleq
\left\{
\left\{\bar{\alpha}_\ell[n]\right\}_{n=0}^{N-1},
\nu_\ell,\lambda_\ell
\right\}_{\ell=0}^{L_h-1}
\label{eq:channel_state}
\end{equation}
collects the physical channel parameters.

The sampled received block is therefore
\begin{equation}
\mathbf y
=
\mathbf H(\bfeps,\mathbf h)\mathbf b+\mathbf n.
\label{eq:observation_model}
\end{equation}
Conditioned on the physical state and available side
information, we assume that $\mathbf n$ is zero-mean
proper complex noise, is uncorrelated with $\mathbf b$
in both the covariance and pseudo-covariance senses,
and has covariance
$\mathbf R_n^{\rm true}\succ\mathbf0$.
The numerical experiments use conditionally Gaussian
noise, $\mathbf n\sim\CN(\mathbf0,\mathbf R_n^{\rm true})$,
while the linear-MMSE results require only these
second-order assumptions.
For compactness, define the true physical state as
\begin{equation}
\bfth
\triangleq
\left(
\bfeps,\mathbf h,\mathbf R_n^{\rm true}
\right).
\label{eq:true_state}
\end{equation}

\subsection{Imperfect Receiver Information}
\label{subsec:side_information}

The equalizer does not know $\bfth$.
Instead, the preceding synchronization and estimation
stages provide
\begin{equation}
\mathsf S
=
\left(
\widehat{\bfeps},
\widehat{\mathbf h},
\widehat{\mathbf R}_n,
\boldsymbol\zeta
\right),
\label{eq:side_information}
\end{equation}
where the hatted quantities are synchronization,
channel, and noise-covariance estimates.
The vector $\boldsymbol\zeta$ contains reliability
descriptors, such as estimated error variances or
pilot-quality indicators.
The conditional law $p(\bfth\mid\mathsf S)$ describes
the remaining physical-state uncertainty without
requiring Gaussian or independent estimation errors.

Physical estimation errors induce the structured
channel mismatch
\begin{equation}
\mathbf E_H(\bfth,\mathsf S)
=
\mathbf H(\bfeps,\mathbf h)
-
\mathbf H(\widehat{\bfeps},\widehat{\mathbf h}),
\label{eq:structured_error}
\end{equation}
rather than an arbitrary matrix perturbation.
We assume that $\mathbf b$ is independent of
$(\bfth,\mathsf S)$, so its conditional moments
equal the unconditional moments defined above.
The side information is obtained from a statistically
separate pilot or preamble.

The formulation also covers a selected linear target
$\mathbf t=\mathbf A\mathbf b$.
The corresponding moments are
$\boldsymbol\mu_t=\mathbf A\boldsymbol\mu_b$,
$\mathbf R_t=\mathbf A\mathbf R_b\mathbf A^H$,
and
$\mathbf C_{y,t}(\bfth)=\mathbf C(\bfth)\mathbf A^H$,
where $\mathbf C(\bfth)$ is defined below.
The same substitutions apply to conditional and
finite-state moments, and the MSE criterion becomes
$\E\|\mathbf t-\widehat{\mathbf t}\|_2^2$.
In the numerical experiments, the target consists
of two phase-derotated dominant Laurent coefficients.

\subsection{Augmented Observation Statistics}
\label{subsec:wl_model}

WL processing augments $\mathbf y\in\mathbb C^N$
with its conjugate:
\begin{equation}
\underline{\mathbf y}
=
\begin{bmatrix}
\mathbf y\\
\mathbf y^*
\end{bmatrix},
\quad
\underline{\mathbf y}\in\mathbb C^{2N}.
\label{eq:augmented_y}
\end{equation}

For a fixed physical state $\bfth$, the conditional
mean of the augmented observation is
\begin{equation}
\boldsymbol\mu_y(\bfth)
=
\E[\underline{\mathbf y}\mid\bfth]
=
\begin{bmatrix}
\mathbf H\boldsymbol\mu_b\\
\mathbf H^*\boldsymbol\mu_b^*
\end{bmatrix},
\label{eq:augmented_mean}
\end{equation}
where $\mathbf H=\mathbf H(\bfeps,\mathbf h)$.

Define the centered observation--symbol
cross-covariance as
\begin{equation}
\mathbf C(\bfth)
=
\E\!\left[
(\underline{\mathbf y}-\boldsymbol\mu_y(\bfth))
(\mathbf b-\boldsymbol\mu_b)^H
\mid\bfth
\right]
=
\begin{bmatrix}
\mathbf H\mathbf R_b\\
\mathbf H^*\mathbf C_b^*
\end{bmatrix},
\label{eq:cross_covariance}
\end{equation}
and the centered augmented observation covariance as
\begin{equation}
\mathbf R_y(\bfth)
=
\operatorname{Cov}(\underline{\mathbf y}\mid\bfth)
=
\begin{bmatrix}
\mathbf P(\bfth)&\mathbf Q(\bfth)\\
\mathbf Q^*(\bfth)&\mathbf P^*(\bfth)
\end{bmatrix},
\label{eq:augmented_covariance}
\end{equation}
with
\begin{equation}
\mathbf P(\bfth)
=
\mathbf H\mathbf R_b\mathbf H^H
+
\mathbf R_n^{\rm true},
\qquad
\mathbf Q(\bfth)
=
\mathbf H\mathbf C_b\mathbf H^T.
\label{eq:PQ_blocks}
\end{equation}

Given $\mathsf S$, the conditional statistics
used by the equalizer are
\begin{align}
\boldsymbol\mu_{\mathsf S}
&=
\E[\underline{\mathbf y}\mid\mathsf S],
\quad
\mathbf R_{\mathsf S}
=
\operatorname{Cov}(\underline{\mathbf y}\mid\mathsf S),
\nonumber\\
\mathbf C_{\mathsf S}
&=
\E\!\left[
(\underline{\mathbf y}-\boldsymbol\mu_{\mathsf S})
(\mathbf b-\boldsymbol\mu_b)^H
\mid\mathsf S
\right].
\label{eq:conditional_statistics}
\end{align}
Uncertainty in the state-dependent mean also contributes
to $\mathbf R_{\mathsf S}$.
Lemma~\ref{lem:posterior_results} gives the corresponding
decomposition.

\subsection{Receiver Classes}
\label{subsec:receiver_classes}

Given $\mathsf S$, consider affine-WL receivers
of the form
\begin{equation}
\begin{split}
\widehat{\mathbf b}
&=
\mathbf a(\mathsf S)
+
\mathbf F(\mathsf S)^H\mathbf y
+
\mathbf G(\mathsf S)^H\mathbf y^*
\\
&=
\mathbf a(\mathsf S)
+
\mathbf W(\mathsf S)^H\underline{\mathbf y},
\end{split}
\label{eq:wl_receiver}
\end{equation}
where
\begin{equation}
\mathbf W(\mathsf S)
=
\begin{bmatrix}
\mathbf F(\mathsf S)\\
\mathbf G(\mathsf S)
\end{bmatrix}.
\end{equation}

Setting $\mathbf G=\mathbf0$ gives complex-linear
processing.
The optimal offset vanishes when the conditional
target and observation means are both zero.
The plug-in (PI) receiver treats the estimated
state as exact.
It forms
$\widehat{\bfth}
=
(\widehat{\bfeps},\widehat{\mathbf h},
\widehat{\mathbf R}_n)$
and, assuming
$\mathbf R_y(\widehat{\bfth})\succ\mathbf0$, uses
\begin{align}
\mathbf W_{\rm PI}(\mathsf S)
&=
\mathbf R_y(\widehat{\bfth})^{-1}
\mathbf C(\widehat{\bfth}),
\\
\mathbf a_{\rm PI}(\mathsf S)
&=
\boldsymbol\mu_b
-
\mathbf W_{\rm PI}(\mathsf S)^H
\boldsymbol\mu_y(\widehat{\bfth}).
\label{eq:plugin_receiver}
\end{align}

This point-estimate design follows the WLTV-MMSE
receiver class considered in~\cite{Darsena2018}.
To assess the benefit of conjugate processing
under synchronization mismatch, we compare it
with a complex-linear plug-in LTV-MMSE receiver
designed from the same estimated state.


\begin{figure*}[!t]
\centering

\definecolor{figink}{RGB}{38,43,51}
\definecolor{figmuted}{RGB}{92,100,112}
\definecolor{statefill}{RGB}{241,246,252}
\definecolor{estfill}{RGB}{250,247,239}
\definecolor{plugfill}{RGB}{252,244,244}
\definecolor{optfill}{RGB}{240,248,243}
\definecolor{oraclefill}{RGB}{242,247,252}
\definecolor{finitefill}{RGB}{247,244,251}

\resizebox{0.86\textwidth}{!}{%
\begin{tikzpicture}[
    x=1cm,
    y=0.70cm,
    font=\small,
    >={Latex[length=2.1mm,width=1.3mm]},
    box/.style={
        draw=figink,
        line width=.62pt,
        rounded corners=2pt,
        align=center,
        inner xsep=9pt,
        inner ysep=6pt,
        minimum height=12mm
    },
    flow/.style={
        ->,
        draw=figink,
        line width=.8pt
    },
    softflow/.style={
        ->,
        draw=figmuted,
        dashed,
        line width=.75pt
    },
    head/.style={
        font=\small\bfseries,
        text=figink,
        anchor=west
    },
    gap/.style={
        font=\scriptsize,
        text=figink,
        align=center
    },
    note/.style={
        font=\scriptsize,
        text=figmuted,
        align=center
    }
]


\node[head] at (-0.1,1.00)
{(a) Information structure};

\node[
    box,
    fill=statefill,
    minimum width=36mm
] (theta) at (0,-.25)
{
    \textbf{True physical state}\\[0.7mm]
    $\bfth=
    (\bfeps,\mathbf h,\mathbf R_n^{\rm true})$
};

\node[
    box,
    fill=estfill,
    minimum width=40mm
] (est) at (6.55,-.25)
{
    \textbf{Synchronization / channel estimator}\\[0.7mm]
    pilot or preamble processing
};

\node[
    box,
    fill=statefill,
    minimum width=43mm
] (S) at (13.10,-.25)
{
    \textbf{Available side information}\\[0.7mm]
    $\mathsf S=
    (\widehat{\bfeps},
     \widehat{\mathbf h},
     \widehat{\mathbf R}_n,
     \boldsymbol\zeta)$
};

\draw[flow] (theta.east) -- (est.west);
\draw[flow] (est.east) -- (S.west);

\node[
    box,
    fill=white,
    minimum width=41mm
] (obs) at (0,-2.45)
{
    \textbf{Sampled augmented observation}\\[0.7mm]
    $\underline{\mathbf y}
    =
    [\,\mathbf y^T\;\mathbf y^H\,]^T$
};

\draw[flow] (theta.south) -- (obs.north);

\draw[figmuted!30,line width=.38pt]
(-2.6,-3.60) -- (15.35,-3.60);


\node[head] at (-0.1,-4.25)
{(b) Receiver hierarchy and performance losses};

\node[
    box,
    fill=plugfill,
    minimum width=34mm
] (pi) at (0,-5.95)
{
    \textbf{Plug-in WL receiver}\\[0.7mm]
    uses $\mathsf S$ through $\widehat{\bfth}$\\[0.7mm]
    $\widehat{\mathbf b}_{\rm PI}$
};

\node[
    box,
    fill=optfill,
    minimum width=41mm
] (opt) at (6.55,-5.95)
{
    \textbf{$\mathsf S$-optimal affine-WL receiver}\\[0.7mm]
    $\mathbf W_\star(\mathsf S)
    =
    \mathbf R_{\mathsf S}^{-1}
    \mathbf C_{\mathsf S}$\\[0.7mm]
    $\widehat{\mathbf b}_{\mathsf S}^{\star}$
};

\node[
    box,
    fill=oraclefill,
    minimum width=34mm
] (oracle) at (13.10,-5.95)
{
    \textbf{Perfect-information oracle}\\[0.7mm]
    uses $(\bfth,\mathsf S)$\\[0.7mm]
    $\widehat{\mathbf b}_{\rm or}$
};

\draw[flow] (pi.east) -- (opt.west);
\draw[flow] (opt.east) -- (oracle.west);

\node[gap] at (3.1,-4.99)
{
    $\mathcal G_{\rm PI}$\\[0.6mm]
    \textit{design loss}
};

\node[gap] at (9.95,-4.99)
{
    $\mathcal G_{\rm info}$\\[0.6mm]
    \textit{information loss}
};


\node[
    box,
    fill=finitefill,
    minimum width=41mm
] (finite) at (6.55,-8.95)
{
    \textbf{Finite-state implementation}\\[0.7mm]
    $Z=\mathcal Q(\mathsf S)
    \in\{1,\ldots,Q\}$\\[0.7mm]
    $\widehat{\mathbf b}_{Z}^{\star}$
};

\draw[softflow] (opt.south) -- (finite.north);

\node[note,anchor=west] at (8.4,-7.55)
{
    compression loss\\[0.6mm]
    $\mathcal J_Z^\star-\mathcal J_{\mathsf S}^\star$
};


\node[
    font=\small\bfseries,
    text=figink
] at (-0.05,-8.35)
{
    Exact orthogonal MSE decomposition
};

\node[
    font=\normalsize,
    text=figink,
    align=center
] at (-0.05,-9.25)
{
    $\displaystyle
    \mathcal J_{\rm PI}
    -
    \mathcal J_{\rm or}^{\star}
    =
    \underbrace{\mathcal G_{\rm PI}}_{\text{design loss}}
    +
    \underbrace{\mathcal G_{\rm info}}_{\text{information loss}}
    $
};

\end{tikzpicture}%
}

\caption{Information structure and affine-WL receiver hierarchy.
The physical state governs the received data and pilot/preamble
observations. Conditional adaptation and revelation of the true
state remove orthogonal design and information losses,
respectively. Compressing side information to $Q$ receiver
states trades adaptation accuracy for stored-filter implementation.}
\label{fig:receiver_hierarchy}
\end{figure*}
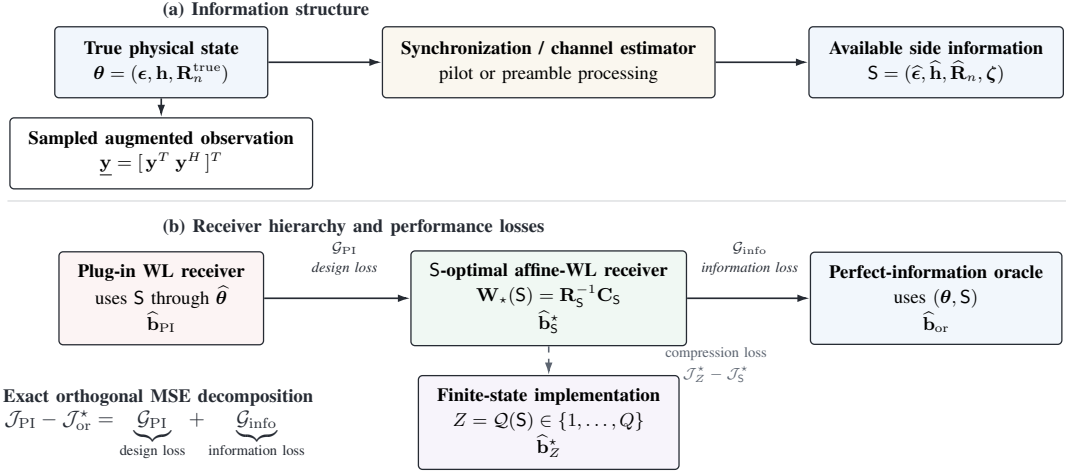


Let
\begin{equation}
\widehat{\mathbf H}
=
\mathbf H(\widehat{\bfeps},\widehat{\mathbf h}),
\qquad
\widehat{\mathbf P}
=
\widehat{\mathbf H}\mathbf R_b\widehat{\mathbf H}^{H}
+
\widehat{\mathbf R}_n,
\label{eq:reference_linear_stats}
\end{equation}
where $\widehat{\mathbf P}$ is the plug-in
covariance of $\mathbf y$.

Assuming $\widehat{\mathbf P}\succ\mathbf0$,
define the complex-linear reference receiver as
\begin{equation}
\widehat{\mathbf b}_{\rm r}
=
\mathbf a_{\rm r}
+
\mathbf F_{\rm r}^{H}\mathbf y,
\label{eq:reference_linear_receiver}
\end{equation}
where
\begin{equation}
\mathbf F_{\rm r}
=
\widehat{\mathbf P}^{-1}
\widehat{\mathbf H}\mathbf R_b,
\qquad
\mathbf a_{\rm r}
=
\boldsymbol\mu_b
-
\mathbf F_{\rm r}^{H}
\widehat{\mathbf H}\boldsymbol\mu_b.
\label{eq:reference_linear_coefficients}
\end{equation}
Thus, $\mathbf G_{\rm r}=\mathbf0$.
In the local viability analysis, both plug-in
receivers are frozen at $\mathsf S=s$ while
the true synchronization state varies.

Let $\mathcal A_{\mathsf S}$ denote the class of
all affine-WL receivers with $\mathsf S$-measurable
coefficients and estimates with finite second moments.
This class allows full use of $\boldsymbol\zeta$
without imposing a plug-in structure.
The oracle class $\mathcal A_{\rm or}$ allows
coefficients measurable with respect to
$\mathsf O=(\bfth,\mathsf S)$, so
$\mathcal A_{\mathsf S}\subseteq\mathcal A_{\rm or}$.
Both classes use the same sampled affine-WL architecture.

\subsection{Problem Formulation}
\label{subsec:problem_formulation}

For an affine-WL receiver
$\mathcal R=(\mathbf a,\mathbf W)\in\mathcal A_{\mathsf S}$,
define
\begin{equation}
\mathcal J(\mathcal R)
=
\E\left[
\left\|
\mathbf b
-
\mathbf a(\mathsf S)
-
\mathbf W(\mathsf S)^H\underline{\mathbf y}
\right\|_2^2
\right].
\label{eq:mse_objective}
\end{equation}

The side-information and perfect-information affine-WL
limits are, respectively,
\begin{align}
\mathcal J_{\mathsf S}^{\star}
&=\inf_{\mathcal R\in\mathcal A_{\mathsf S}}\mathcal J(\mathcal R),
\label{eq:linear_limit}\\
\mathcal J_{\rm or}^{\star}
&=\inf_{\mathcal R\in\mathcal A_{\rm or}}\mathcal J(\mathcal R).
\label{eq:oracle_limit}
\end{align}
The information and plug-in design losses are
\begin{align}
\mathcal G_{\rm info}
&=\mathcal J_{\mathsf S}^{\star}-\mathcal J_{\rm or}^{\star},
\label{eq:information_gap}\\
\mathcal G_{\rm PI}
&=\mathcal J_{\rm PI}-\mathcal J_{\mathsf S}^{\star},
\label{eq:plugin_gap}
\end{align}
where $\mathcal J_{\rm PI}=\mathcal J(\mathbf a_{\rm PI},\mathbf W_{\rm PI})$.
Consequently,
\begin{equation}
\mathcal J_{\rm PI}
-
\mathcal J_{\rm or}^{\star}
=
\mathcal G_{\rm PI}
+
\mathcal G_{\rm info}.
\label{eq:gap_decomposition}
\end{equation}
Section~\ref{sec:theoretical_results} gives the
exact orthogonal MSE interpretation of these losses.

For the synchronization-robustness analysis,
we freeze both plug-in receivers at the same
estimated state $\widehat{\bfth}(s)$.
Let
$\boldsymbol\xi
\triangleq
(\mathbf h,\mathbf R_n^{\rm true})$
collect the physical quantities held fixed
while the residual synchronization vector
$\bfeps$ varies.
For a fixed affine receiver $\mathcal R$, define
\begin{equation}
M_{\mathcal R}(\bfeps;\boldsymbol\xi)
=
\E\left[
\|\mathbf b-\widehat{\mathbf b}\|_2^2
\mid\bfeps,\boldsymbol\xi
\right].
\label{eq:pointwise_mse}
\end{equation}
The synchronization viability region of the
plug-in WLTV receiver is
\begin{equation}
\mathcal V_s(\boldsymbol\xi)
=
\left\{
\bfeps:
M_{\mathcal R_{\rm PI}}(\bfeps;\boldsymbol\xi)
<
M_{\mathcal R_{\rm r}}(\bfeps;\boldsymbol\xi)
\right\},
\label{eq:viability_region}
\end{equation}
where $\mathcal R_{\rm r}$ is the complex-linear
reference receiver.
Thus, $\mathcal V_s(\boldsymbol\xi)$ identifies
the synchronization states for which the frozen
WLTV receiver achieves lower MSE than its
complex-linear counterpart.
Figure~\ref{fig:receiver_hierarchy} summarizes
the information structure and receiver hierarchy.

\section{Performance Limits and Receiver Design}
\label{sec:theoretical_results}

We first establish the conditional affine-WL statistics
and quantify the local synchronization tolerance of
frozen plug-in receivers. We then characterize the
side-information and oracle performance limits and
use these results to design finite-state receivers.
Proofs are provided in the supplementary appendices.

\subsection{Statistical Preliminaries}
\label{subsec:results_preliminaries}

Using the augmented statistics of
Section~\ref{subsec:wl_model}, we make the following
assumptions.

\emph{A1) Conditional data statistics:}
almost surely,
$\E[\mathbf b\mid\bfth,\mathsf S]=\boldsymbol\mu_b$,
with conditional covariance $\mathbf R_b$ and
pseudo-covariance $\mathbf C_b$.

\emph{A2) Data-block noise:}
conditioned on $(\bfth,\mathsf S)$, the noise is
zero-mean and proper, has covariance
$\mathbf R_n^{\rm true}$, and is uncorrelated with
the centered data in both the covariance and
pseudo-covariance senses.

\emph{A3) Uniform noise floor:}
there exists $\sigma_0^2>0$ such that
$\mathbf R_n^{\rm true}\succeq\sigma_0^2\mathbf I$
almost surely.

\emph{A4) Finite second moments:}
$\E\|\mathbf b\|_2^2<\infty$ and
$\E\|\underline{\mathbf y}\|_2^2<\infty$.

We impose no Gaussianity or independence assumptions
on the timing, frequency, channel, or noise-covariance
estimation errors. Their joint uncertainty is described
by $p(\bfth\mid\mathsf S)$.

\begin{lemma}
\label{lem:posterior_results}
Under A1--A4, the observation statistics conditioned
on the available side information satisfy
$\boldsymbol\mu_{\mathsf S}
=\E[\boldsymbol\mu_y(\bfth)\mid\mathsf S]$
and
$\mathbf C_{\mathsf S}
=\E[\mathbf C(\bfth)\mid\mathsf S]$.
The observation covariance decomposes as
\begin{equation}
\mathbf R_{\mathsf S}
=
\E[\mathbf R_y(\bfth)\mid\mathsf S]
+
\operatorname{Cov}
\left(\boldsymbol\mu_y(\bfth)\mid\mathsf S\right),
\label{eq:results_total_cov}
\end{equation}
and
$\mathbf R_{\mathsf S}\succeq
\sigma_0^2\mathbf I_{2N}$ almost surely.
\end{lemma}

\emph{Proof:}
See Appendix~\ref{app:proof_lemma_posterior}.

Thus, the conditional covariance accounts for both
the average physical-state covariance and variation
in the state-dependent mean. Evaluating the statistics
at a single state estimate generally does not capture
these effects.

\subsection{Conditional Affine-WL Estimation}
\label{subsec:conditional_affine_results}

Let $U$ denote side information satisfying
$\E[\mathbf b\mid U]=\boldsymbol\mu_b$ and
$\operatorname{Cov}(\mathbf b\mid U)=\mathbf R_b$
almost surely, and consider
$\widehat{\mathbf b}
=\mathbf a(U)+\mathbf W(U)^H\underline{\mathbf y}$.
Define
$\boldsymbol\mu_U=\E[\underline{\mathbf y}\mid U]$,
$\mathbf R_U
=\operatorname{Cov}(\underline{\mathbf y}\mid U)$,
and
$\mathbf C_U
=\E[(\underline{\mathbf y}-\boldsymbol\mu_U)
(\mathbf b-\boldsymbol\mu_b)^H\mid U]$.
The following identity applies to the available
information $U=\mathsf S$, its finite-state
representation $U=Z=\mathcal Q(\mathsf S)$,
and the oracle information
$U=\mathsf O=(\bfth,\mathsf S)$.

\begin{lemma}[Conditional affine-WL MSE]
\label{lem:affine_identity_results}
For any admissible offset $\mathbf a(U)$ and
filter $\mathbf W(U)$,
\begin{align}
\mathcal J(\mathbf a,\mathbf W\mid U)
&\triangleq
\E\!\left[
\|\mathbf b-\mathbf a-\mathbf W^H\underline{\mathbf y}\|_2^2
\mid U
\right]
\label{eq:results_affine_identity}\\
&=
\tr(\mathbf R_b)
+
\|\boldsymbol\mu_b-\mathbf a-\mathbf W^H\boldsymbol\mu_U\|_2^2
\nonumber\\
&\quad
-2\Re\!\left\{\tr(\mathbf W^H\mathbf C_U)\right\}
+\tr(\mathbf W^H\mathbf R_U\mathbf W)
\nonumber
\end{align}
almost surely. For a fixed $\mathbf W(U)$,
the MSE-minimizing offset is
\[
\mathbf a(U)
=
\boldsymbol\mu_b-\mathbf W(U)^H\boldsymbol\mu_U.
\]
\end{lemma}

\emph{Proof:}
See Appendix~\ref{app:proof_lemma_affine}.

\subsection{Local Timing--Frequency Viability of the Plug-In Receiver}
\label{subsec:viability_results}

We first quantify the synchronization mismatch
over which a frozen plug-in WLTV receiver retains
its MSE advantage over the complex-linear reference.
Fix $\mathsf S=s$, the receivers
$\mathcal R_{\rm PI}$ and $\mathcal R_{\rm r}$,
the design point $\bfeps_0=\widehat{\bfeps}(s)$,
and the physical quantities
$\boldsymbol\xi=(\mathbf h,\mathbf R_n^{\rm true})$.
Only $\bfeps=(\epsilon_\tau,\epsilon_\nu)^T$ varies.

Write
$\widehat{\bfth}
=(\widehat{\bfeps},\widehat{\mathbf h},
\widehat{\mathbf R}_n)$
for the estimated state, and let
$\widehat{\boldsymbol\mu}_y
=\boldsymbol\mu_y(\widehat{\bfth})$,
$\widehat{\mathbf R}_y
=\mathbf R_y(\widehat{\bfth})$, and
$\widehat{\mathbf C}=\mathbf C(\widehat{\bfth})$
denote the corresponding statistics.
Provided $\widehat{\mathbf R}_y\succ\mathbf0$,
the plug-in WLTV receiver uses
\begin{equation}
\mathbf W_{\rm PI}
=
\widehat{\mathbf R}_y^{-1}\widehat{\mathbf C},
\qquad
\mathbf a_{\rm PI}
=
\boldsymbol\mu_b
-\mathbf W_{\rm PI}^H\widehat{\boldsymbol\mu}_y.
\label{eq:results_plugin_receiver}
\end{equation}

The complex-linear reference $\mathcal R_{\rm r}$
in \eqref{eq:reference_linear_receiver}--%
\eqref{eq:reference_linear_coefficients}
is designed from the same $\widehat{\bfth}$ and
held fixed in the same way. The comparison therefore
isolates the benefit of WL processing under a common
synchronization mismatch.

The channel depends on $\bfeps$ through the pulse
shift and common residual-frequency phase
in \eqref{eq:true_channel_matrix}.
For differentiable effective pulses, define
$\eta=T_s/T$ and
$\mathbf D
=\operatorname{diag}(t_0/T_s,t_0/T_s+1,\ldots,
t_0/T_s+N-1)$.
Using subscripts $\tau$ and $\nu$ for derivatives
with respect to $\epsilon_\tau$ and $\epsilon_\nu$,
respectively, the first channel derivatives are
\begin{align}
[\mathbf H_\tau]_{n,k}
&=
-T\sum_{\ell=0}^{L_h-1}
\bar{\alpha}_\ell[n]\,
e^{j2\pi(t_0+nT_s)(\nu_\ell+\epsilon_\nu/T)}
\nonumber\\
&\quad\times
\phi_k'(nT_s-\lambda_\ell-T\epsilon_\tau),
\nonumber\\
\mathbf H_\nu
&=
j2\pi\eta\mathbf D\mathbf H.
\label{eq:results_Htau}
\end{align}
The corresponding derivatives of
$\boldsymbol\mu_y$, $\mathbf C$, and $\mathbf R_y$,
together with the first- and second-order MSE
derivatives used below, are given in
Appendix~\ref{app:derivatives}.

For a fixed affine receiver
$\mathcal R=(\mathbf a,\mathbf W)$, define
$\mathbf m_{\mathcal R}
=\boldsymbol\mu_b-\mathbf a-\mathbf W^H\boldsymbol\mu_y$.
Its pointwise MSE is
\begin{align}
M_{\mathcal R}(\bfeps;\boldsymbol\xi)
&=
\tr(\mathbf R_b)
+\|\mathbf m_{\mathcal R}\|_2^2
-2\Re\{\tr(\mathbf W^H\mathbf C)\}
\nonumber\\
&\quad+\tr(\mathbf W^H\mathbf R_y\mathbf W).
\label{eq:results_pointwise_MSE}
\end{align}
Because the receiver coefficients are fixed,
the gradient and Hessian with respect to $\bfeps$
depend only on derivatives of the state-dependent
statistics.

Define the relative MSE as
\[
D_s(\bfeps;\boldsymbol\xi)
=
M_{\mathcal R_{\rm PI}}(\bfeps;\boldsymbol\xi)
-
M_{\mathcal R_{\rm r}}(\bfeps;\boldsymbol\xi).
\]
The exact viability region is
$\mathcal V_s(\boldsymbol\xi)
=\{\bfeps:D_s(\bfeps;\boldsymbol\xi)<0\}$.
At $\bfeps_0$, define the nominal margin
$\Gamma_s=-D_s(\bfeps_0;\boldsymbol\xi)$,
sensitivity
$\mathbf g_s
=\nabla_{\bfeps}D_s(\bfeps_0;\boldsymbol\xi)$,
and curvature
$\mathbf B_s
=\nabla_{\bfeps}^2D_s(\bfeps_0;\boldsymbol\xi)$.
These derivatives follow from
\eqref{eq:results_first_mse_derivative}--%
\eqref{eq:results_second_mse_derivative}
by subtracting the reference-receiver derivatives
from those of the plug-in WLTV receiver.

\begin{theorem}[Local relative-MSE viability]
\label{thm:viability_results}
Fix $\mathsf S=s$ and $\boldsymbol\xi$.
Suppose $D_s(\bfeps;\boldsymbol\xi)$ is three times
continuously differentiable in a neighborhood of
$\bfeps_0$, and $\Gamma_s>0$.
For $\boldsymbol\delta=\bfeps-\bfeps_0$, define
\[
q_s(\boldsymbol\delta)
\triangleq
-\Gamma_s
+\mathbf g_s^T\boldsymbol\delta
+\frac12\boldsymbol\delta^T\mathbf B_s\boldsymbol\delta.
\]
For sufficiently small $\boldsymbol\delta$,
\begin{equation}
D_s(\bfeps_0+\boldsymbol\delta;\boldsymbol\xi)
=
q_s(\boldsymbol\delta)+R_3(\boldsymbol\delta),
\label{eq:results_Taylor}
\end{equation}
where $R_3$ is the third-order remainder and
$D_s(\bfeps_0;\boldsymbol\xi)=-\Gamma_s<0$.

Let
$\mathcal B_r
=\{\boldsymbol\delta:\|\boldsymbol\delta\|_2\le r\}$
for $r>0$ such that $\bfeps_0+\mathcal B_r$
lies within this neighborhood. Define
\[
L_3(r)
\triangleq
\sup_{\boldsymbol\delta\in\mathcal B_r}
\left\|
\nabla_{\bfeps}^{3}
D_s(\bfeps_0+\boldsymbol\delta;\boldsymbol\xi)
\right\|_{\rm op},
\]
where
\[
\|\nabla_{\bfeps}^{3}D_s(\mathbf x;\boldsymbol\xi)\|_{\rm op}
\triangleq
\sup_{\substack{\|\mathbf u\|_2=\|\mathbf v\|_2=1\\
\|\mathbf w\|_2=1}}
\left|
\nabla_{\bfeps}^{3}D_s(\mathbf x;\boldsymbol\xi)
[\mathbf u,\mathbf v,\mathbf w]
\right|.
\]
For finite $L_3(r)$, Taylor's theorem gives
$|R_3(\boldsymbol\delta)|
\le L_3(r)\|\boldsymbol\delta\|_2^3/6$
on $\mathcal B_r$.

Define the inner and outer regions
\begin{align}
\mathcal V_{\rm in}(r)
&=
\left\{
\bfeps_0+\boldsymbol\delta:
\|\boldsymbol\delta\|_2\le r,
\right.
\nonumber\\
&\hspace{1.3cm}\left.
q_s(\boldsymbol\delta)
+\frac{L_3(r)}{6}\|\boldsymbol\delta\|_2^3<0
\right\},
\label{eq:results_Vin}\\[0.5em]
\mathcal V_{\rm out}(r)
&=
\left\{
\bfeps_0+\boldsymbol\delta:
\|\boldsymbol\delta\|_2\le r,
\right.
\nonumber\\
&\hspace{1.3cm}\left.
q_s(\boldsymbol\delta)
-\frac{L_3(r)}{6}\|\boldsymbol\delta\|_2^3<0
\right\}.
\label{eq:results_Vout}
\end{align}
Then
\begin{equation}
\mathcal V_{\rm in}(r)
\subseteq
\mathcal V_s(\boldsymbol\xi)
\cap(\bfeps_0+\mathcal B_r)
\subseteq
\mathcal V_{\rm out}(r).
\label{eq:results_viability_sandwich}
\end{equation}
The inner region certifies retained WL advantage,
while the outer region contains every viable point
within radius $r$.
\end{theorem}

\emph{Proof:}
See Appendix~\ref{app:proof_viability}.

Neglecting the remainder gives the approximate
local boundary
\begin{equation}
\mathbf g_s^T\boldsymbol\delta
+\frac12\boldsymbol\delta^T\mathbf B_s\boldsymbol\delta
=
\Gamma_s,
\label{eq:results_quadratic_boundary}
\end{equation}
where $\Gamma_s$, $\mathbf g_s$, and $\mathbf B_s$
describe the nominal WL advantage, sensitivity,
and curvature, respectively.
If $\mathbf B_s\succ\mathbf0$, this boundary is
an ellipse centered at
$-\mathbf B_s^{-1}\mathbf g_s$
in $\boldsymbol\delta$-coordinates.
The certified regions additionally account for
the remainder.

A sufficient condition for every point in
$\bfeps_0+\mathcal B_r$ to remain viable is
\begin{equation}
-\Gamma_s+\|\mathbf g_s\|_2 r
+\frac12\max\{\lambda_{\max}(\mathbf B_s),0\}r^2
+\frac{L_3(r)}{6}r^3<0.
\end{equation}
This condition follows by uniformly bounding the
linear and quadratic terms on $\mathcal B_r$.
It provides an analysis certificate for fixed
$\boldsymbol\xi$ and a valid upper bound on $L_3(r)$;
using it as an online test would require information
beyond $\mathsf S$.

\subsection{Best Receiver Using the Available Side Information}
\label{subsec:sideinfo_results}

We next characterize the best affine-WL receiver
using $\mathsf S$ without imposing a plug-in structure.
This benchmark identifies the loss that receiver
redesign can remove without improving the front-end
information.
By Lemma~\ref{lem:posterior_results},
$\mathbf R_{\mathsf S}\succeq
\sigma_0^2\mathbf I_{2N}$ almost surely.

\begin{theorem}[Side-information-optimal affine-WL receiver]
\label{thm:sideinfo_results}
Under A1--A4, the minimum-MSE affine-WL receiver
based on $\mathsf S$ is unique up to almost-sure
equality and is given by
\begin{equation}
\begin{aligned}
\widehat{\mathbf b}_{\mathsf S}^{\star}
&=
\mathbf a_\star(\mathsf S)
+\mathbf W_\star(\mathsf S)^H\underline{\mathbf y},
\\
\mathbf W_\star(\mathsf S)
&=
\mathbf R_{\mathsf S}^{-1}\mathbf C_{\mathsf S},
\end{aligned}
\label{eq:results_Wstar}
\end{equation}
with
$\mathbf a_\star(\mathsf S)
=\boldsymbol\mu_b
-\mathbf W_\star(\mathsf S)^H\boldsymbol\mu_{\mathsf S}$.
Its minimum unconditional MSE is
\begin{equation}
\mathcal J_{\mathsf S}^{\star}
=
\tr(\mathbf R_b)
-
\E\!\left[
\tr\!\left(
\mathbf C_{\mathsf S}^H
\mathbf R_{\mathsf S}^{-1}
\mathbf C_{\mathsf S}
\right)
\right].
\label{eq:results_JSstar}
\end{equation}

For any other admissible affine-WL receiver
$(\mathbf a(\mathsf S),\mathbf W(\mathsf S))$
using the same side information, the excess MSE is
\begin{align}
\mathcal J(\mathbf a,\mathbf W)
-\mathcal J_{\mathsf S}^{\star}
&=
\E\Big[
\|\boldsymbol\mu_b-\mathbf a
-\mathbf W^H\boldsymbol\mu_{\mathsf S}\|_2^2
\label{eq:results_excess_risk}\\
&\;+
\tr\!\left(
(\mathbf W-\mathbf W_\star)^H
\mathbf R_{\mathsf S}
(\mathbf W-\mathbf W_\star)
\right)
\Big].
\nonumber
\end{align}
Both terms are nonnegative and quantify the
penalties due to mean mismatch and filter mismatch,
respectively.
\end{theorem}

\emph{Proof:}
See Appendix~\ref{app:proof_sideinfo}.

The classical affine/WL-MMSE
projection~\cite{Picinbono1995,SchreierScharf2010,Kay1993}
therefore gives an architecture-specific performance
limit using conditional moments over the physical
states compatible with $\mathsf S$.

The distinction from posterior-mean parameter
substitution is already visible when
$\boldsymbol\mu_b=\mathbf0$.
Define
$\overline{\mathbf H}_{\mathsf S}
=\E[\mathbf H\mid\mathsf S]$
and
$\Delta\mathbf H
=\mathbf H-\overline{\mathbf H}_{\mathsf S}$.
The ordinary covariance block satisfies
\begin{align}
\E[\mathbf P(\bfth)\mid\mathsf S]
&=
\overline{\mathbf H}_{\mathsf S}\mathbf R_b
\overline{\mathbf H}_{\mathsf S}^{H}
+
\E[\Delta\mathbf H\mathbf R_b
\Delta\mathbf H^H\mid\mathsf S]
\nonumber\\
&\quad+
\E[\mathbf R_n^{\rm true}\mid\mathsf S],
\end{align}
while the complementary covariance block is
\begin{align}
\E[\mathbf Q(\bfth)\mid\mathsf S]
&=
\overline{\mathbf H}_{\mathsf S}\mathbf C_b
\overline{\mathbf H}_{\mathsf S}^{T}
\nonumber\\
&\quad+
\E[\Delta\mathbf H\mathbf C_b
\Delta\mathbf H^T\mid\mathsf S].
\end{align}
Thus, even substituting the conditional mean channel
omits uncertainty-dependent covariance terms.
Moreover, because $\mathbf H(\bfeps,\mathbf h)$
is nonlinear in the physical parameters,
\[
\mathbf H\!\left(
\E[\bfeps\mid\mathsf S],
\E[\mathbf h\mid\mathsf S]
\right)
\ne
\E[\mathbf H(\bfeps,\mathbf h)\mid\mathsf S]
\]
in general.

For the plug-in receiver,
\eqref{eq:results_excess_risk} gives
\begin{align}
\mathcal G_{\rm PI}
&=
\E\Big[
\|\mathbf W_{\rm PI}^H
(\widehat{\boldsymbol\mu}_y-\boldsymbol\mu_{\mathsf S})
\|_2^2
\label{eq:results_GPI}\\
&\quad+
\tr\!\left(
(\mathbf W_{\rm PI}-\mathbf W_\star)^H
\mathbf R_{\mathsf S}
(\mathbf W_{\rm PI}-\mathbf W_\star)
\right)
\Big]
\ge0.
\nonumber
\end{align}
The first term penalizes the use of the point-estimate
mean in place of the conditional mean; the second
penalizes filter mismatch.
Hence, $\mathcal G_{\rm PI}$ is a pure
\emph{design loss}: it can be reduced while retaining
the same front-end information.

\subsection{Oracle Performance and Design--Information Decomposition}
\label{subsec:oracle_results}

Let $\mathsf O=(\bfth,\mathsf S)$ additionally reveal
the physical state.
Because $\mathsf O$ includes $\mathsf S$, every
$\mathsf S$-based receiver is oracle-admissible.
Under A1--A2, the observation statistics conditioned
on $\mathsf O$ are $\boldsymbol\mu_y(\bfth)$,
$\mathbf R_y(\bfth)$, and $\mathbf C(\bfth)$.
The oracle provides a perfect-information benchmark
within the same sampled affine-WL architecture.

\begin{theorem}[Oracle receiver and orthogonal loss decomposition]
\label{thm:oracle_results}
Under A1--A4, the minimum-MSE affine-WL receiver
based on $\mathsf O=(\bfth,\mathsf S)$ is
\begin{equation}
\begin{aligned}
\widehat{\mathbf b}_{\rm or}
&=
\mathbf a_{\rm or}(\bfth)
+
\mathbf W_{\rm or}(\bfth)^H\underline{\mathbf y},
\\
\mathbf W_{\rm or}(\bfth)
&=
\mathbf R_y(\bfth)^{-1}\mathbf C(\bfth),
\end{aligned}
\label{eq:results_oracle}
\end{equation}
where
$\mathbf a_{\rm or}(\bfth)
=\boldsymbol\mu_b
-\mathbf W_{\rm or}(\bfth)^H\boldsymbol\mu_y(\bfth)$.
Its average MSE is
\begin{equation}
\mathcal J_{\rm or}^{\star}
=
\tr(\mathbf R_b)
-
\E\!\left[
\tr\!\left(
\mathbf C(\bfth)^H
\mathbf R_y(\bfth)^{-1}
\mathbf C(\bfth)
\right)
\right].
\label{eq:results_Jor}
\end{equation}

The information gap
$\mathcal G_{\rm info}
=\mathcal J_{\mathsf S}^{\star}
-\mathcal J_{\rm or}^{\star}$
satisfies
\begin{equation}
\mathcal G_{\rm info}
=
\E\!\left[
\|\widehat{\mathbf b}_{\mathsf S}^{\star}
-\widehat{\mathbf b}_{\rm or}\|_2^2
\right]
\ge0.
\label{eq:results_Ginfo_projection}
\end{equation}
Thus, the loss due to incomplete physical-state
information equals the mean-square distance
between the side-information-optimal and oracle
affine-WL estimates.

The total degradation of the plug-in receiver
relative to the oracle satisfies
\begin{equation}
\mathcal J_{\rm PI}
-\mathcal J_{\rm or}^{\star}
=
\mathcal G_{\rm PI}+\mathcal G_{\rm info},
\label{eq:results_total_gap}
\end{equation}
and the corresponding estimator corrections are
orthogonal in the unconditional mean-square sense:
\begin{equation}
\E\!\left[
(\widehat{\mathbf b}_{\rm PI}
-\widehat{\mathbf b}_{\mathsf S}^{\star})^H
(\widehat{\mathbf b}_{\mathsf S}^{\star}
-\widehat{\mathbf b}_{\rm or})
\right]
=
0.
\label{eq:results_orthogonality}
\end{equation}
\end{theorem}

\emph{Proof:}
See Appendix~\ref{app:proof_oracle}.

Equation~\eqref{eq:results_orthogonality} separates
the benefit of receiver redesign using unchanged
side information from the benefit of revealing
the physical state.

Equivalently,
\begin{align}
\mathcal G_{\rm info}
&=
\E\Big[
\|\mathbf W_\star^H
(\boldsymbol\mu_y(\bfth)-\boldsymbol\mu_{\mathsf S})
\|_2^2
\label{eq:results_Ginfo_explicit}\\
&\quad+
\tr\!\left(
(\mathbf W_\star-\mathbf W_{\rm or})^H
\mathbf R_y(\bfth)
(\mathbf W_\star-\mathbf W_{\rm or})
\right)
\Big].
\nonumber
\end{align}
Information loss therefore arises from uncertainty
in the state-dependent observation mean, the optimal
WL coefficient matrix, or both.
A dominant $\mathcal G_{\rm PI}$ indicates greater
potential benefit from equalizer redesign, whereas
a dominant $\mathcal G_{\rm info}$ indicates greater
potential benefit from more informative front-end
estimation.

\begin{corollary}[Zero information loss]
\label{cor:zero_information_results}
The information gap vanishes if and only if
$\widehat{\mathbf b}_{\mathsf S}^{\star}
=\widehat{\mathbf b}_{\rm or}$
almost surely.
\end{corollary}

\emph{Proof:}
See Appendix~\ref{app:proof_zero_info}.

Zero information loss therefore does not require
exact recovery of every physical parameter.
It requires only that the remaining state ambiguity
leave the optimal affine-WL estimate unchanged,
providing a receiver-oriented notion of sufficient
front-end information.

\subsection{Finite-State Approximation}
\label{subsec:finite_state_results}

Continuous adaptation of
$(\mathbf a_\star(\mathsf S),\mathbf W_\star(\mathsf S))$
may require repeated evaluation of conditional
statistics and synthesis of filter coefficients.
A finite-state implementation replaces these online
operations with selection from a stored filter bank.
Context-tree and clustering turbo equalizers compress
decoder side information~\cite{Kim2013Context,Kim2014Clustering}.
Here, $Z=\mathcal Q(\mathsf S)\in\{1,\ldots,Q\}$
compresses uncertain physical-state information
and selects one stored receiver from
$\{(\mathbf a_j,\mathbf W_j)\}_{j=1}^{Q}$.

For implementation, the state decision can use
a lower-dimensional routing feature
$\boldsymbol\psi(\mathsf S)$:
\begin{equation}
Z
=
\widetilde{\mathcal Q}
\!\left(\boldsymbol\psi(\mathsf S)\right),
\qquad
Z\in\{1,\ldots,Q\}.
\label{eq:results_feature_quantizer}
\end{equation}
Choosing $\boldsymbol\psi(\mathsf S)=\mathsf S$
recovers the unrestricted formulation.
Practical features may include synchronization
estimates, channel and noise quality indicators,
and reliability descriptors in $\boldsymbol\zeta$.

For every state $j$ with $\Pr(Z=j)>0$, define
$\boldsymbol\mu_j
=\E[\underline{\mathbf y}\mid Z=j]$,
$\mathbf R_j
=\operatorname{Cov}(\underline{\mathbf y}\mid Z=j)$,
and
\[
\mathbf C_j
=
\E[
(\underline{\mathbf y}-\boldsymbol\mu_j)
(\mathbf b-\boldsymbol\mu_b)^H
\mid Z=j].
\]

Because $Z$ is a deterministic function of
$\mathsf S$, A1 and the tower property preserve
the target moments when conditioning on $Z$.
Under A1--A4, the conditional covariance decomposition
also gives
$\mathbf R_j\succeq\sigma_0^2\mathbf I_{2N}$.
Thus, every state of positive probability admits
a unique affine-WL MMSE solution.

\begin{theorem}[Optimal finite-state affine-WL receiver and routing-information monotonicity]
\label{thm:finite_state_results}
For a fixed quantizer $\mathcal Q$, the unique
minimum-MSE affine-WL receiver for each state $j$
with $\Pr(Z=j)>0$ is
\begin{equation}
\mathbf W_j^\star
=
\mathbf R_j^{-1}\mathbf C_j,
\qquad
\mathbf a_j^\star
=
\boldsymbol\mu_b
-
(\mathbf W_j^\star)^H\boldsymbol\mu_j.
\label{eq:results_finite_state_filter}
\end{equation}
Let $\widehat{\mathbf b}_Z^\star$ denote the estimate
obtained by selecting these coefficients according
to $Z=\mathcal Q(\mathsf S)$. Then
\begin{equation}
\mathcal J_Z^\star-\mathcal J_{\mathsf S}^{\star}
=
\E\!\left[
\left\|
\widehat{\mathbf b}_Z^\star
-\widehat{\mathbf b}_{\mathsf S}^{\star}
\right\|_2^2
\right]
\ge0.
\label{eq:results_finite_state_penalty}
\end{equation}
Thus, the exact MSE cost of finite-state adaptation
equals the mean-square distance between the optimal
finite-state and continuously adapted estimates.

More generally, let $\mathfrak Q_Q(\boldsymbol\psi)$
denote the class of deterministic $Q$-state quantizers
that depend on $\mathsf S$ only through
$\boldsymbol\psi(\mathsf S)$, and define
\begin{equation}
\mathcal J_{Q,\boldsymbol\psi}^{\star}
\triangleq
\inf_{\mathcal Q\in\mathfrak Q_Q(\boldsymbol\psi)}
\mathcal J_{\mathcal Q}^{\star},
\label{eq:results_feature_optimum}
\end{equation}
where $\mathcal J_{\mathcal Q}^{\star}$ is the MSE
obtained using the optimal coefficients
in \eqref{eq:results_finite_state_filter}
for the partition induced by $\mathcal Q$.
Then
\begin{equation}
\mathcal J_{\mathsf S}^{\star}
\le
\mathcal J_{Q,\boldsymbol\psi}^{\star}.
\label{eq:results_feature_lower_bound}
\end{equation}
If
$\boldsymbol\psi_1(\mathsf S)
=g(\boldsymbol\psi_2(\mathsf S))$
almost surely for some measurable $g$, then
\begin{equation}
\mathcal J_{Q,\boldsymbol\psi_2}^{\star}
\le
\mathcal J_{Q,\boldsymbol\psi_1}^{\star}.
\label{eq:results_feature_monotonicity}
\end{equation}
Hence, at a fixed state budget $Q$, adding available
reliability information to the routing feature
cannot increase the best achievable population MSE.
\end{theorem}

\emph{Proof:}
See Appendix~\ref{app:proof_finite_state}.

Nearby parameter values need not yield similar
equalizers. We therefore measure the cost of
sharing a filter directly through excess MSE.
For $\mathsf S=s$ and a candidate stored receiver
$(\mathbf a,\mathbf W)$, define
\begin{equation}
\begin{aligned}
d(s;\mathbf a,\mathbf W)
&=
\left\|
\boldsymbol\mu_b-\mathbf a
-\mathbf W^H\boldsymbol\mu_{\mathsf S}(s)
\right\|_2^2
\\
&\quad+
\tr\!\Bigl(
(\mathbf W-\mathbf W_\star(s))^H
\mathbf R_{\mathsf S}(s)
\\
&\hspace{4em}\times
(\mathbf W-\mathbf W_\star(s))
\Bigr).
\end{aligned}
\label{eq:results_distortion}
\end{equation}
By Theorem~\ref{thm:sideinfo_results}, this distortion
is the conditional excess MSE relative to continuous
adaptation.

\begin{proposition}[Generalized Lloyd conditions]
\label{prop:lloyd_results}
For a fixed codebook
$\{(\mathbf a_j,\mathbf W_j)\}_{j=1}^{Q}$,
an unrestricted MSE-minimizing assignment satisfies
\begin{equation}
\mathcal Q^\star(s)
\in
\arg\min_{1\le j\le Q}
d(s;\mathbf a_j,\mathbf W_j)
\label{eq:results_Lloyd_assignment}
\end{equation}
for almost every $s$ under the distribution of
$\mathsf S$.

For a fixed partition of the side-information space,
the MSE-minimizing codeword for each state of
positive probability is the conditional affine-WL
MMSE receiver in
\eqref{eq:results_finite_state_filter}.
\end{proposition}

\emph{Proof:}
See Appendix~\ref{app:proof_lloyd}.

This proposition specializes the classical Lloyd/LBG
alternating-minimization
principle~\cite{Lloyd1982,Linde1980}
to the equalization distortion
in \eqref{eq:results_distortion}.
Alternating minimum-distortion assignments and
conditional-MMSE codeword updates produces
nonincreasing population MSE values.
These values therefore converge, although global
optimality is not guaranteed.
Finite-sample designs require validation.

The bank is designed offline using a model or
training ensemble.
To avoid the online conditional-statistic evaluations
that exact distortion routing may require, our
numerical implementation uses threshold routing with
$\boldsymbol\psi(\mathsf S)
=[\widehat\epsilon_\tau,\widehat\epsilon_\nu,
\boldsymbol\zeta_{\rm sync}^T]^T$.
Nonuniform thresholds may also initialize Lloyd
refinement without changing the observation
architecture.

\section{Numerical Evaluation}
\label{sec:numerical_evaluation}

We evaluate finite-state adaptation against the
point-estimate (PI) and posterior-mean (PM)
substitution receivers, a matched-budget
representative-state bank, a numerical
continuous-adaptation reference, and a
matched-state/oracle reference.
These comparisons quantify the gains from
uncertainty-aware coefficient design and the
cost of finite-state implementation.
Theorem~\ref{thm:viability_results} provides
the analytical robustness certificate.

\subsection{Simulation Setup}
\label{subsec:num_setup}

The binary GMSK link operates at $200$~kbit/s
with symbol duration $T=5~\mu$s, carrier frequency
$5.06$~GHz, modulation index $1/2$,
Gaussian-filter bandwidth--time product $BT=0.30$,
memory $L=3$, and eight samples per symbol.
A four-symbol observation window contains
$32$ complex samples.
The known reference channel has delays
$(0,3.5,7)~\mu$s, Doppler shifts
$(1520,-700,250)$~Hz, phases $(0,0.8,-1.1)$,
and powers $(\kappa,0.75,0.25)/(\kappa+1)$,
where $\kappa=10^{2.9}$.
Unless otherwise stated, $t_0=60T=300~\mu$s;
phase rotations use $t_0+nT_s$, while pulse
arguments use $nT_s$.

The dominant-component implementation retains
$18$ Laurent coefficients, indexed by
$m=-9,\ldots,8$.
Writing $b_m=j^{m+1}u_m$, with equiprobable
$u_m\in\{-1,+1\}$, allows the phase rotations
to be absorbed into the equivalent real channel
matrix $\mathbf B\in\mathbb R^{64\times18}$.
We estimate $\mathbf t=[u_0,u_{-1}]^T$ from
$\mathbf y_{\rm R}
=[\Re\{\mathbf y\}^T\;\Im\{\mathbf y\}^T]^T$,
an equivalent real representation of augmented
WL processing.
Each filter has $64\times2$ real coefficients,
and bit decisions use
$\operatorname{sign}(\widehat u_0\widehat u_{-1})$.
MSE is averaged over the two target coordinates.
Filter design and analytical MSE calculations
use the dominant component; empirical full-CPM
MSE and ABER use the full waveform.
With $\gamma=10^{(E_b/N_0)_{\rm dB}/10}$,
the complex white-noise variance is $8/\gamma$,
and the variance per real coordinate is
$n_r=8/(2\gamma)$.

The true residual timing and common-frequency
states, denoted here by $\tau$ and $f$,
are independent zero-mean Gaussians with
\begin{equation}
\operatorname{std}(\tau/T)=0.20,
\qquad
\operatorname{std}(f)=400~\mathrm{Hz}.
\label{eq:num_true_state_spread}
\end{equation}
Independently of these states and the transmitted
data, synchronization-estimation errors are
zero-mean Gaussian with standard deviations
$(\sigma_\tau,\sigma_f)$.
The reported standard deviations equal the
actual values except in the miscalibration
experiment.
The high-uncertainty setting is
\begin{equation}
\sigma_\tau/T=0.25,
\qquad
\sigma_f=500~\mathrm{Hz}.
\label{eq:num_high_uncertainty}
\end{equation}
PI synthesizes its filter at
$(\widehat\tau,\widehat f)$.
PM first computes the Gaussian posterior-mean
state using the same reported uncertainty
available to the bank, then substitutes this
state into the filter design.
The bank routes using features derived from
\begin{equation}
\boldsymbol\psi_2
=
(\widehat\tau,\widehat f,\sigma_\tau,\sigma_f).
\label{eq:num_rich_feature}
\end{equation}

Offline development uses $4500$ training states
and $1500$ independent validation states.
Their error standard deviations are drawn
independently from log-uniform distributions
on $[0.02,0.35]T$ and $[25,600]$~Hz.
For each $Q$, candidate empirical-quantile
partitions use timing and frequency bin counts
whose product is $Q$, in either raw-estimate
or posterior-mean coordinates.
Conditional moments are obtained by averaging
model moments over training states, with
pseudo-symbols averaged analytically.
At $E_b/N_0=8$~dB, validation MSE selects the
conditional-bank partition, after which the
bank is refitted using all $6000$ development
states.
At each $E_b/N_0$, the corresponding noise
variance is inserted into the same conditional
moments.
The evaluated design uses this restricted
threshold search; generalized Lloyd refinement
and unrestricted population optimization are
not performed.

The SNR and state-budget tests use $100$
synchronization-state clusters, $800$ symbol
contexts per cluster, and $800$ paired
cluster-bootstrap resamples.
The channel, age, calibration, and
continuous-reference tests use independent
seeds, $256$ clusters, $400$ contexts per
cluster at $E_b/N_0=8$~dB, and $2000$
bootstrap resamples.
Receivers share waveform and noise realizations,
and test labels are never used to select
partitions.
Intervals are pointwise, except for the
channel-ensemble aggregate, which resamples
channels within each power-ratio stratum.
The supplied code, seeds, channel parameters,
and cluster outputs support reproduction
of these tests.

\begin{figure*}[!t]
\centering
\begin{minipage}[t]{0.49\textwidth}
    \centering
    \includegraphics[width=\linewidth]{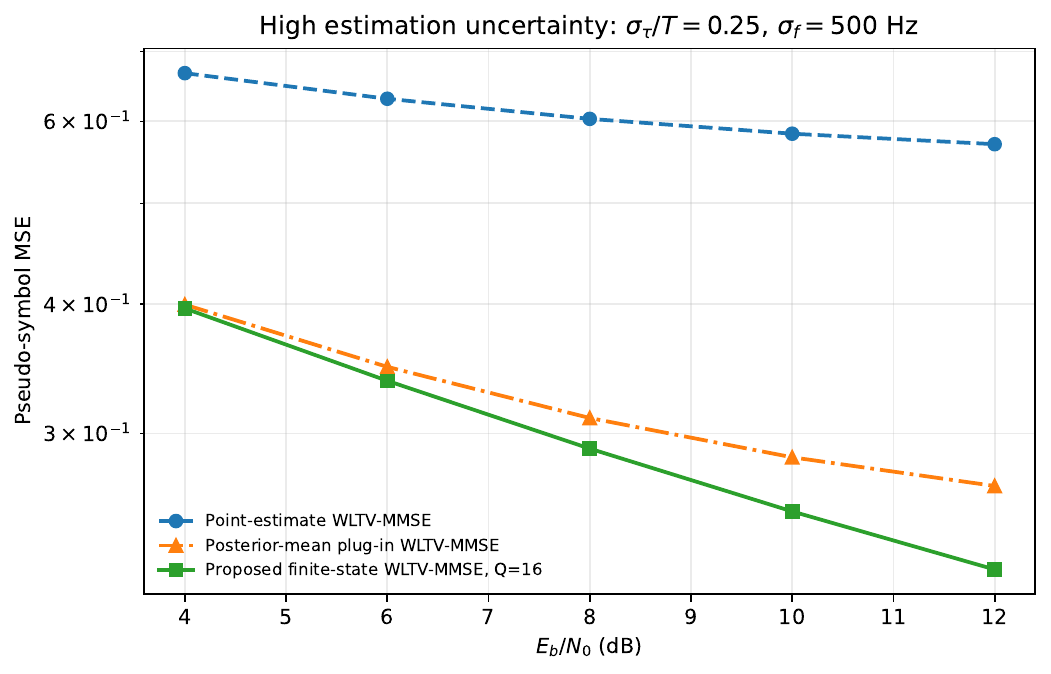}
    \par\smallskip
    \textbf{(a)}
\end{minipage}
\hfill
\begin{minipage}[t]{0.49\textwidth}
    \centering
    \includegraphics[width=\linewidth]{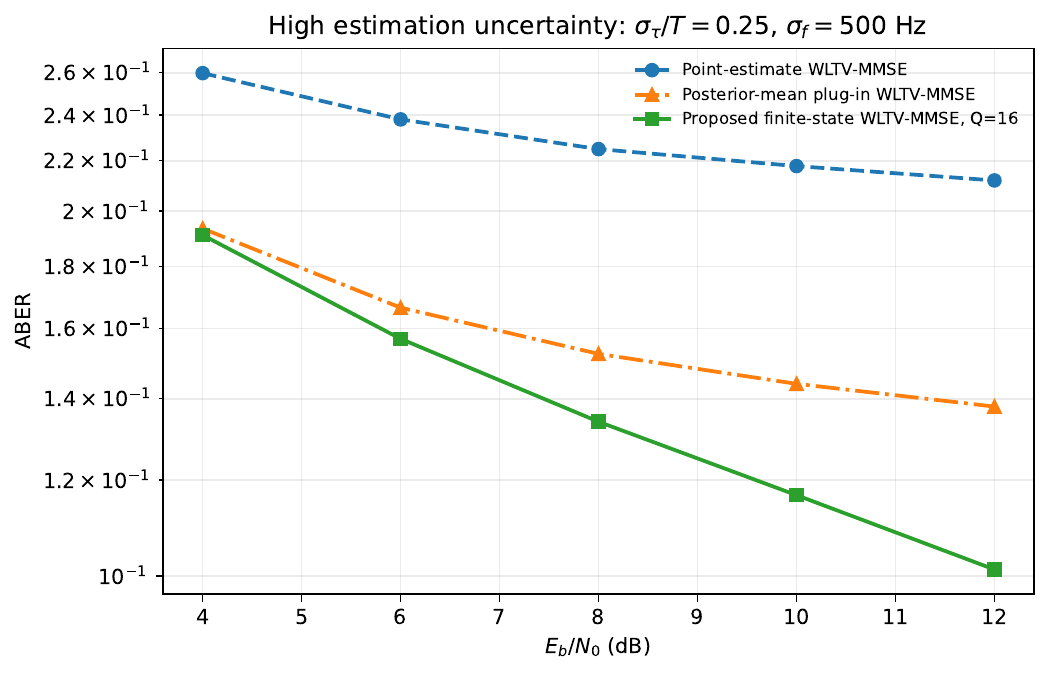}
    \par\smallskip
    \textbf{(b)}
\end{minipage}
\caption{Performance under high synchronization-estimation
uncertainty
$(\sigma_\tau/T,\sigma_f)=(0.25,500~\mathrm{Hz})$.
(a) Pseudo-symbol MSE.
(b) Full-CPM ABER.
The posterior-mean substitution receiver uses the same
uncertainty information as the proposed method.
The comparison quantifies the additional empirical gain
of the selected conditional-moment filter bank over
this baseline.}
\label{fig:t4_snr}
\end{figure*}

\subsection{Reference-Channel Performance and State Budget}
\label{subsec:num_snr}
\label{subsec:num_routing_validation}

Figure~\ref{fig:t4_snr} shows reference-channel
performance versus $E_b/N_0$.
At $E_b/N_0=8$~dB, the bank reduces pseudo-symbol
MSE by $51.9\%$ relative to PI, empirical
full-CPM MSE from $0.60402$ to $0.29175$,
and ABER from $0.22493$ to $0.13404$.
PM attains $0.31053$, $0.31149$, and $0.15246$
in these three metrics, respectively.
The bank therefore provides additional
reductions of $6.6\%$, $6.3\%$, and $12.1\%$.
Paired $95\%$ intervals for the absolute
PM-minus-bank reductions are
$[0.01009,0.03114]$, $[0.00913,0.03005]$,
and $[0.01130,0.02653]$, respectively.
At $E_b/N_0=12$~dB, the corresponding relative
reductions reach $16.9\%$, $16.5\%$, and $26.6\%$.
These results refer to the SNR-test ensemble;
the robustness and continuous-reference tests
below use independent ensembles.

The PI baseline implements conventional
WLTV-MMSE state substitution within the common
sampled GMSK model by treating the estimated
physical state as exact.
The aeronautical CPM/WLTV framework
in~\cite{Darsena2018} provides the closest
application-specific precedent.
The reported gains quantify uncertainty-aware
adaptation relative to this controlled PI
implementation.

\begin{figure}[!t]
\centering
\includegraphics[width=0.98\columnwidth]{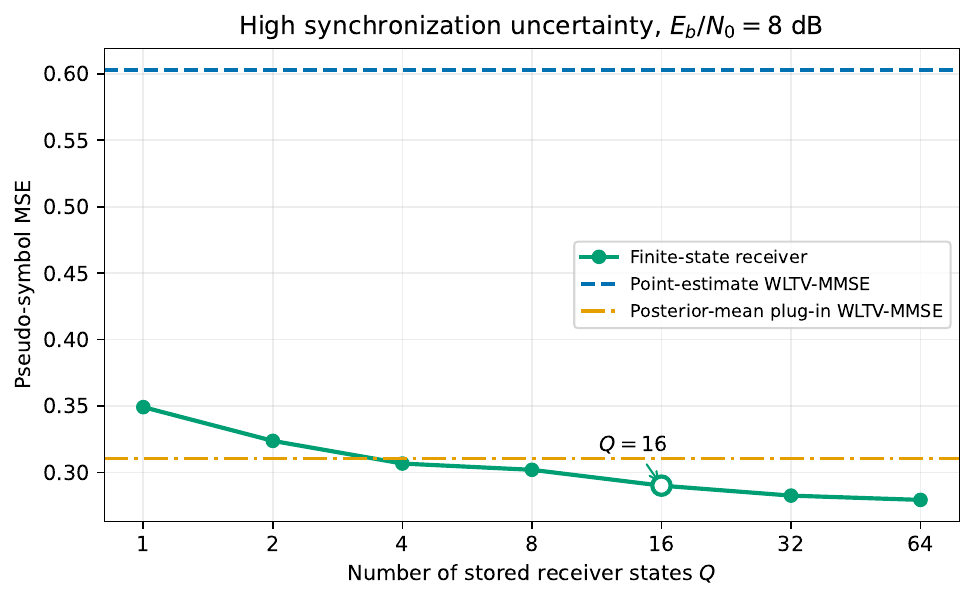}
\caption{Pseudo-symbol MSE versus the number $Q$ of stored
receiver states at $E_b/N_0=8$~dB under high
synchronization-estimation uncertainty
$(\sigma_\tau/T,\sigma_f)=(0.25,500~\mathrm{Hz})$.
The horizontal lines show the point-estimate and
posterior-mean substitution WLTV-MMSE baselines.
The highlighted $Q=16$ point is the operating point
used in the SNR and uncertainty-plane experiments.}
\label{fig:q_budget}
\end{figure}

Figure~\ref{fig:q_budget} evaluates
$Q\in\{1,2,4,8,16,32,64\}$ at $E_b/N_0=8$~dB
under \eqref{eq:num_high_uncertainty}.
PI and PM attain pseudo-symbol MSEs of
$0.60303$ and $0.31053$, while the $Q=16$
and $Q=64$ banks attain $0.29013$ and $0.27939$.
Thus, $Q=16$ retains $96.7\%$ of the improvement
achieved by $Q=64$ relative to PI, using one
quarter as many stored filter coefficients.
Increasing $Q$ from $16$ to $64$ provides
a further $3.7\%$ MSE reduction.
This comparison quantifies the storage tradeoff
within the evaluated bank family; the gap to
continuous adaptation is assessed separately
below.

At $Q=16$, validation selects one timing bin
and sixteen posterior-frequency bins, with
routing coordinate
$400^2\widehat f/(400^2+\sigma_f^2)$ in Hz.
This selected router uses frequency confidence
but not timing confidence.
The richer routing family reduces validation
MSE from $0.15130$ to $0.13902$ relative to
raw-estimate routing.
Timing resolution first appears at $Q=32$
in this sweep.

To isolate the coefficient-design gain, we
compare two banks with the same $Q=16$
partition, thresholds, routing rule, storage,
and online cost.
The shared partition is selected using
conditional-bank validation MSE; only the
stored coefficients differ.
The comparison bank replaces each
conditional-moment filter with a conventional
WLTV-MMSE filter synthesized at a representative
point state.
For cell $j$, this state is the empirical mean
of the posterior-mean timing and frequency
coordinates of the development samples routed
to that cell.
Neither test labels nor true test states
are used.

On the same high-uncertainty reference test
at $E_b/N_0=8$~dB, the representative-state
bank attains pseudo-symbol MSE $0.32286$
and full-waveform ABER $0.15884$.
The conditional-moment bank attains $0.29013$
and $0.13404$, corresponding to reductions
of $10.1\%$ and $15.6\%$, respectively.
Paired cluster-bootstrap $95\%$ intervals
for the absolute improvements are
$[0.02541,0.04065]$ and $[0.01919,0.03075]$.
Under the shared partition, this comparison
isolates the benefit of designing coefficients
from the conditional uncertainty distribution
rather than from representative point states.

\subsection{Timing--Frequency Uncertainty Plane}
\label{subsec:num_uncertainty_plane}

\begin{figure*}[t]
\centering
\begin{minipage}[t]{0.49\textwidth}
    \centering
    \includegraphics[width=\linewidth]{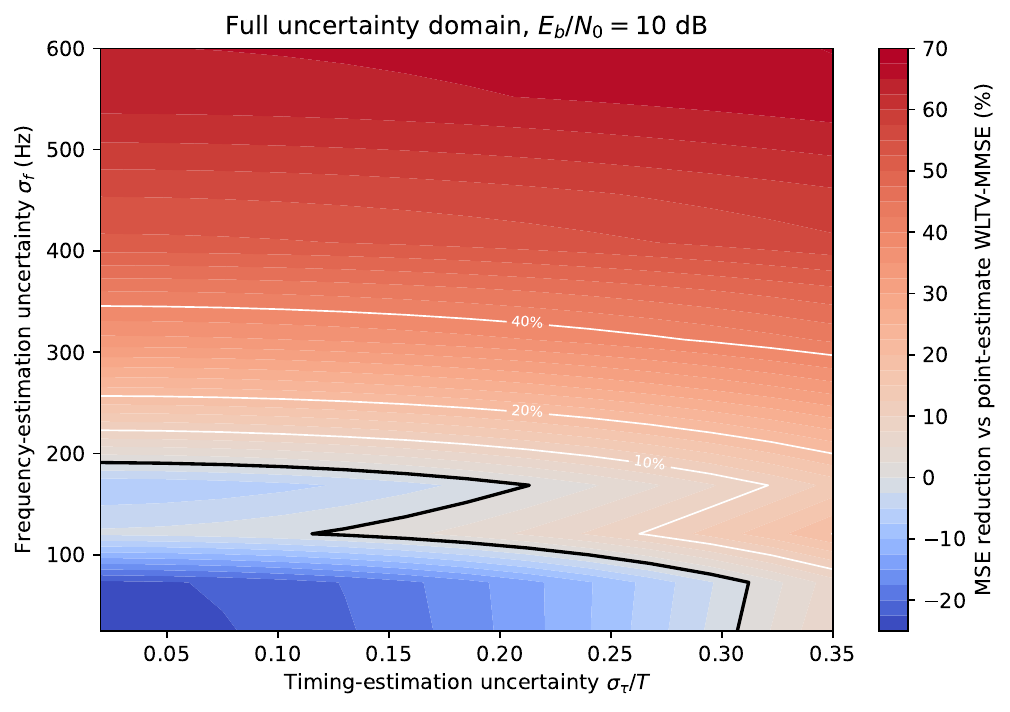}
    \par\smallskip
    \textbf{(a)}
\end{minipage}
\hfill
\begin{minipage}[t]{0.49\textwidth}
    \centering
    \includegraphics[width=\linewidth]{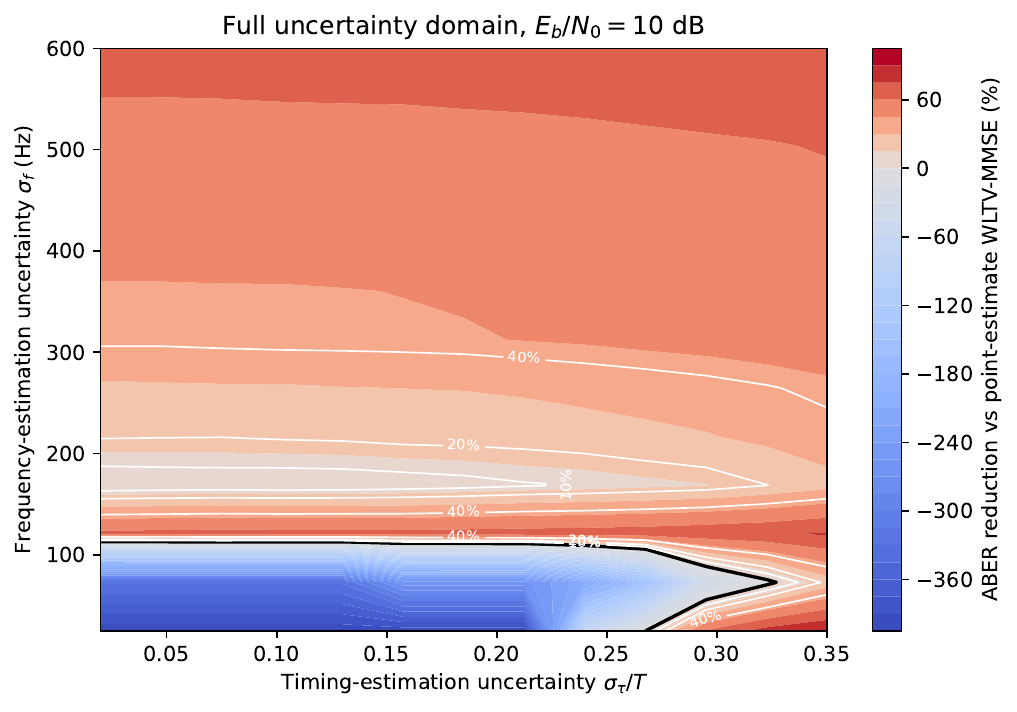}
    \par\smallskip
    \textbf{(b)}
\end{minipage}
\caption{Synchronization-estimation uncertainty over the
evaluated domain at $E_b/N_0=10$~dB.
(a) Relative MSE reduction and
(b) relative full-CPM ABER reduction of the proposed
$Q=16$ receiver with respect to point-estimate WLTV-MMSE.
The black contour denotes zero gain; warm regions favor
the proposed receiver and cool regions favor
point-estimate substitution.}
\label{fig:t4_uncertainty_maps}
\end{figure*}

At $E_b/N_0=10$~dB, a $13\times13$ grid covers
\begin{equation}
0.02\le\sigma_\tau/T\le0.35,
\qquad
25\le\sigma_f\le600~\mathrm{Hz},
\label{eq:num_uncertainty_range}
\end{equation}
with the physical-state distribution unchanged.
Each point uses $24$ clusters and $500$
contexts per cluster.
True states, standardized estimation errors,
data, and noise are reused across grid points.
For the ABER surface ratios only, the error
rate is regularized to $(e+1/2)/(12000+1)$,
where $e$ is the error count.

Figure~\ref{fig:t4_uncertainty_maps} shows the
complete evaluated domain, including regions
of negative gain.
MSE and ABER improve over PI at $79.9\%$
and $87.0\%$ of sampled points, respectively,
and both improve at $79.3\%$.
In the region $\sigma_\tau/T\ge0.08$ and
$\sigma_f\ge200$~Hz, both metrics improve
at every sampled point.
Within this region, mean reductions are
$47.9\%$ and $49.5\%$ relative to PI, and
$5.7\%$ and $16.0\%$ relative to PM,
for MSE and ABER, respectively.
Large negative relative gains at low uncertainty
can reflect small baseline error rates.

\subsection{Channel, Age, and Calibration Sensitivity}
\label{subsec:num_robustness}

We generate eighteen synthetic three-path
channels, with six channels for each
dominant-to-remaining-path power ratio
$K_c\in\{6,15,29\}$~dB.
Powers remain proportional to
$(10^{K_c/10},0.75,0.25)$.
The two nonzero delays are sorted independent
uniform draws on $[0,7]~\mu$s, path Doppler
shifts are independent uniform draws on
$[-1520,1520]$~Hz, and non-dominant phases
are independent uniform draws on $[-\pi,\pi]$,
with the dominant phase set to zero.
Each channel is known, remains fixed during
its test, and has its own trained and validated
bank.
This experiment evaluates performance across
channel conditions using individually fitted
banks.

\begin{table}[t]
\footnotesize
\centering
\caption{Channel-ensemble reductions relative to PM at
$E_b/N_0=8$~dB, $t_0=300~\mu$s, and high uncertainty.
Each stratum contains six channels; gains are computed
from ratios of averaged errors.}
\label{tab:channels}
\begin{tabular}{c|rr}
\hline
$K_c$ (dB)&MSE reduction (\%)&ABER reduction (\%)\\
\hline
$6$&$5.08$&$11.79$\\
$15$&$5.30$&$13.63$\\
$29$&$5.52$&$13.43$\\
All eighteen&$5.30$&$12.93$\\
\hline
\end{tabular}
\end{table}

Both metrics improve for each of the eighteen
channels; Table~\ref{tab:channels} summarizes
the gains by power-ratio stratum.
Aggregate pseudo-symbol MSE decreases from
$0.28399$ for PM to $0.26895$ for the bank,
while ABER decreases from $0.12824$ to $0.11166$.
Stratified channel-bootstrap $95\%$ intervals
for the relative reductions are
$[4.53,6.06]\%$ and $[11.04,14.45]\%$,
respectively.
Full-CPM MSE decreases from $0.28586$ to
$0.27096$, a reduction of $5.21\%$.
These intervals describe the synthetic channel
ensemble conditional on the fitted banks.

\begin{figure*}[t]
\centering
\includegraphics[width=\textwidth]{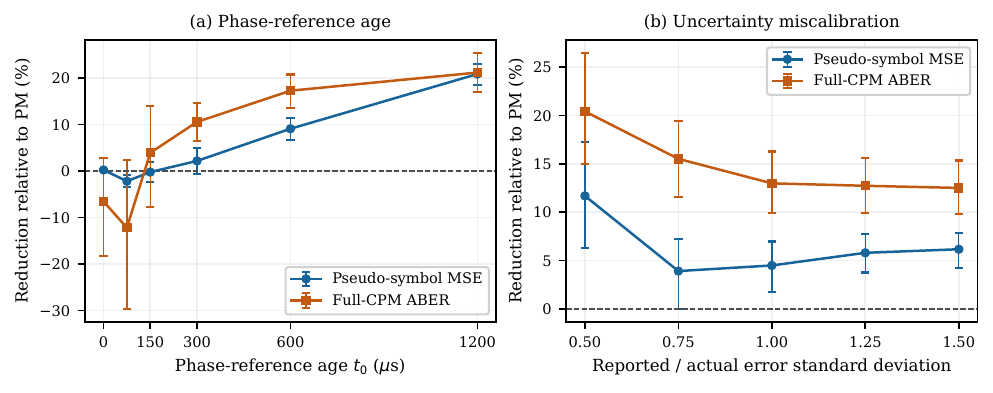}
\caption{Robustness tests at $E_b/N_0=8$~dB relative to PM.
Error bars are paired cluster-bootstrap $95\%$ intervals.
(a) Reference-channel phase-reference-age sweep,
with a separately fitted bank at each age and common
test draws.
(b) Fixed reference-channel bank with both reported
error standard deviations multiplied by $c$, while
actual errors remain unchanged.
Positive values favor the bank; all predefined
conditions are shown.}
\label{fig:robustness}
\end{figure*}

For the reference channel, we next vary
$t_0\in\{0,75,150,300,600,1200\}~\mu$s,
refitting the bank at each phase-reference age.
Figure~\ref{fig:robustness}(a) shows how the
benefit of finite-state adaptation depends
on this age.
At $75~\mu$s, MSE and ABER increase by
$2.2\%$ and $12.2\%$, respectively, relative to PM.
At $150~\mu$s, the MSE difference is
inconclusive.
The independent test ensemble at $300~\mu$s
gives a $2.2\%$ MSE reduction, with a confidence
interval that includes zero.
At $600$ and $1200~\mu$s, relative MSE
reductions increase to $9.1\%$ and $20.9\%$,
respectively, although absolute errors also increase.
At $1200~\mu$s, the bank MSE is $1.08684$,
exceeding the zero-estimate MSE of one.
These results show that relative gains must
be interpreted alongside absolute error levels
as the phase reference ages.

For calibration sensitivity, the actual error
standard deviations remain
$(0.25T,500~\mathrm{Hz})$, while the receiver
is given $(c\,0.25T,c\,500~\mathrm{Hz})$,
with $c\in\{0.5,0.75,1,1.25,1.5\}$.
The bank and thresholds remain fixed;
only the reported-confidence routing feature
changes.
PM uses the same reported values.
Figure~\ref{fig:robustness}(b) shows positive
mean MSE and ABER gains at every tested $c$.
MSE reductions range from $3.9\%$ to $11.7\%$,
and ABER reductions range from $12.5\%$
to $20.4\%$.
The MSE interval at $c=0.75$ includes zero.
For the two largest values of $c$, the reported
frequency uncertainty also exceeds the training
range.
These results quantify sensitivity to the tested
multiplicative miscalibration.

\subsection{Numerical Continuous-Adaptation Reference}
\label{subsec:num_continuous}

On an independent $256$-cluster reference-channel
test at $E_b/N_0=8$~dB, we numerically integrate
the Gaussian posterior of $(\tau/T,f)$ conditioned
on each complete side-information realization.
Tensor Gauss--Hermite quadrature with $40$ nodes
per coordinate provides the conditional real
moments used to construct the affine-WL reference
of Theorem~\ref{thm:sideinfo_results}.
Using $24$ nodes per coordinate changes the
per-realization conditional risk by at most
$1.64\times10^{-7}$.
The resulting filters incur at most
$5.27\times10^{-12}$ additional risk when
evaluated using the finer quadrature.
These numerical convergence checks do not
constitute rigorous integration-error bounds.

\begin{table}[t]
\footnotesize
\centering
\caption{Independent reference test.
Conditional model MSE averages posterior-integrated
risks over side information; full-CPM quantities
use sampled physical states and waveforms.}
\label{tab:continuous}
\begin{tabular}{l|rrr}
\hline
Receiver&Conditional&Full-CPM&Full-CPM\\
&model MSE&MSE&ABER\\
\hline
PI&$0.62485$&$0.59612$&$0.23425$\\
PM&$0.27223$&$0.25671$&$0.09896$\\
Bank, $Q=16$&$0.25767$&$0.24397$&$0.08511$\\
Continuous&$0.23590$&$0.22657$&$0.06670$\\
Matched/oracle&$0.07804$&$0.08104$&$0.00084$\\
\hline
\end{tabular}
\end{table}

Table~\ref{tab:continuous} distinguishes
posterior-integrated model risks from empirical
full-waveform results.
These quantities can differ because of both
the waveform model and finite sampling.
The matched-state/oracle receiver uses true
synchronization; its integrated risk averages
the matched-state MMSE over each posterior.
For the dominant-component model, the
conditional-risk averages give
\begin{align*}
\mathcal G_{\rm PI}
&\simeq0.38895,
&
\mathcal G_{\rm info}
&\simeq0.15786,
\\
\mathcal J_{\rm bank}
-\mathcal J_{\mathsf S}^{\star}
&\simeq0.02176.
\end{align*}
All values are per target coordinate.
The bank's excess risk has a paired $95\%$
interval of $[0.02006,0.02369]$, equals
$9.2\%$ of the continuous-reference risk,
and agrees with the quadratic excess-risk
identity to $1.23\times10^{-15}$.

This excess risk includes the implemented
bank's partition restrictions and
moment-estimation error; it is not the minimum
population compression penalty.
The continuous reference attains full-CPM
MSE and ABER of $0.22657$ and $0.06670$,
compared with $0.24397$ and $0.08511$
for the bank.
The comparison quantifies the remaining
performance gain available through continuous
adaptation, whose online computational cost
must also be considered.

\section{Discussion and Design Implications}
\label{sec:discussion}

For the same target, observation model, and side information,
an admissible implemented bank satisfies
\begin{equation}
\mathcal J_{\rm PI}-\mathcal J_{\rm bank}
=
\mathcal G_{\rm PI}
-
\bigl(
\mathcal J_{\rm bank}-\mathcal J_{\mathsf S}^{\star}
\bigr).
\label{eq:bank_gain_condition}
\end{equation}
Thus, strict MSE improvement occurs precisely when
$\mathcal J_{\rm bank}-\mathcal J_{\mathsf S}^{\star}
<\mathcal G_{\rm PI}$.
In practical terms, the benefit of accounting for
physical-state uncertainty must exceed the loss
incurred by the implemented finite-state approximation.

The reference experiment at $E_b/N_0=8$~dB demonstrates
this regime: the sixteen-state bank reduces pseudo-symbol
MSE by $51.9\%$ relative to point-estimate WLTV-MMSE and
by $6.6\%$ relative to posterior-mean substitution.
In the uncertainty plane sampled at $E_b/N_0=10$~dB,
both MSE and ABER improve over point-estimate
adaptation at every tested point satisfying
$\sigma_\tau/T\ge0.08$ and $\sigma_f\ge200$~Hz.

For the implemented zero-mean receiver, applying
a $64\times2$ filter requires $128$ real
multiplications and $126$ additions per two-target
window.
A $Q=16$ bank stores $2048$ real coefficients
($16$~KiB in 64-bit precision), plus $15$ thresholds.
After feature computation, binary-search frequency
routing requires at most four comparisons.
These counts exclude common sampling,
synchronization, and bit decisions.
Changes in the channel, phase-reference age, or
noise level require appropriate bank updates.

\begin{table}[t]
\footnotesize
\centering
\caption{Single-thread Python CPU timings per window.
Medians and 10th--90th percentiles are computed over
$41$ randomly interleaved blocks of $128$ calls.
Front-end processing and bit decisions are excluded.}
\label{tab:timing}
\begin{tabular}{l|rr}
\hline
Operation&Median ($\mu$s)&10th--90th ($\mu$s)\\
\hline
Routing only&$2.95$&$2.86$--$3.27$\\
Filter only&$1.77$&$1.72$--$2.62$\\
Routing + filter&$4.30$&$4.03$--$5.05$\\
PI, dense&$395.75$&$376.47$--$479.25$\\
PM, dense&$396.26$&$378.78$--$456.02$\\
PI, 18-column form&$376.28$&$356.78$--$449.85$\\
PM, 18-column form&$374.03$&$360.80$--$437.61$\\
\hline
\end{tabular}
\end{table}

Table~\ref{tab:timing} uses one thread on an AMD EPYC 9V74
CPU with Python~3.12.14, NumPy~2.3.5, and SciPy~1.17.0,
including Python call and loop overhead. Plug-in timings
include channel/moment formation, Cholesky factorization,
and filter application. Besides the dense $64\times64$ solve,
an equivalent implementation exploits $\mathbf B$'s $18$ columns:
\[
\mathbf W
=
\mathbf B
(\mathbf B^T\mathbf B+n_r\mathbf I_{18})^{-1}
\mathbf E,
\]
where $\mathbf E$ selects the two target coordinates.
The implementations agree within $1.7\times10^{-15}$ in output.
Median routing-plus-application time is $4.30~\mu$s, versus
$374.03~\mu$s for the faster measured PM implementation:
an approximately $87$-fold reduction with PM synthesis
performed every window.

A conditional-moment refresh using $6000$ development states
and the fixed selected partition takes a median $1.64$~s
over seven runs; partition reselection is additional.
Updating only the noise variance and refactoring sixteen
stored moment matrices takes $0.42$~ms. Against PM synthesis
once per window, these costs are amortized after approximately
$4430$ and $2$ windows, respectively. These are software
measurements, not embedded-device latency or energy guarantees.
Less frequent plug-in adaptation reduces the savings;
rapidly changing conditions can make refresh cost dominant.

The adaptation complements low-complexity WL and banded
structures~\cite{Darsena2018,Chayot2018Band,chae2023low}
and partitioned turbo equalization~\cite{Kim2013Context,Kim2014Clustering}.
The matched-budget experiment isolates conditional-moment
design gains with partition, storage, and routing fixed;
it does not reproduce complete published receivers.
Unknown or drifting channels, non-Gaussian estimation errors,
training variability, and embedded implementation require
further evaluation. The architecture is most useful when
uncertainty-aware design improves accuracy, memory accommodates
the bank, and conditions persist long enough to amortize refresh.

\section{Conclusion}
\label{sec:conclusion}

We developed finite-state, uncertainty-aware WL equalization
for aeronautical GMSK links with limited onboard computation
and memory. Synchronization estimates and reliability
information select stored filters computed from conditional
physical-state moments. Exact excess-MSE identities, a
design--information loss decomposition, and local mismatch
certificates characterize adaptation accuracy and robustness.

On the reference channel at $E_b/N_0=8$~dB under high
synchronization-estimation uncertainty, the $Q=16$ bank
reduces pseudo-symbol MSE by $51.9\%$ relative to
point-estimate WLTV-MMSE and by $6.6\%$ relative to
posterior-mean substitution. 

The bank stores $2048$ real coefficients and replaces
repeated online synthesis with filter selection and application.
It improves on point-estimate adaptation when its excess risk
relative to continuous conditional adaptation is smaller than
the plug-in design loss. The measured online and refresh costs
support its use when side information enables accurate filter
selection and operating conditions persist long enough
to amortize bank updates.

\section*{Acknowledgment}
OpenAI's ChatGPT assisted with language editing, organization,
and LaTeX presentation of the manuscript, and with clarifying
explanations and intermediate derivations in the supplementary
appendices. The authors retain responsibility for the final
content and its verification.

\bibliographystyle{IEEEtran}
\bibliography{references_submission_final}

\clearpage

\appendices
\noindent\textbf{Supplementary material: derivations and proofs}\par

\section{Synchronization-Derivative Formulas}
\label{app:derivatives}

Throughout this appendix, the receiver coefficients
$(\mathbf a,\mathbf W)$ and the physical quantities
$\boldsymbol\xi=(\mathbf h,\mathbf R_n^{\rm true})$
are held fixed.
Subscripts $q,r\in\{\tau,\nu\}$ denote partial
derivatives with respect to the normalized
synchronization coordinates $\epsilon_\tau$
and $\epsilon_\nu$, respectively.
All state-dependent quantities are evaluated at
the same $(\bfeps,\boldsymbol\xi)$.
The formulas below assume that the effective
pulses possess the required derivatives.

Recall that $\eta=T_s/T$ and
\[
\mathbf D
=
\operatorname{diag}
(t_0/T_s,t_0/T_s+1,\ldots,t_0/T_s+N-1).
\]
The first-order channel derivatives are given
in \eqref{eq:results_Htau}.
The corresponding mean and cross-covariance
derivatives are
\[
\boldsymbol\mu_{y,q}
=
\begin{bmatrix}
\mathbf H_q\boldsymbol\mu_b\\
\mathbf H_q^*\boldsymbol\mu_b^*
\end{bmatrix},
\qquad
\mathbf C_q
=
\begin{bmatrix}
\mathbf H_q\mathbf R_b\\
\mathbf H_q^*\mathbf C_b^*
\end{bmatrix}.
\]

Writing
\[
\mathbf R_y
=
\begin{bmatrix}
\mathbf P&\mathbf Q\\
\mathbf Q^*&\mathbf P^*
\end{bmatrix},
\]
with
$\mathbf P=\mathbf H\mathbf R_b\mathbf H^H
+\mathbf R_n^{\rm true}$
and
$\mathbf Q=\mathbf H\mathbf C_b\mathbf H^T$,
we obtain
\[
\begin{aligned}
\mathbf P_q
&=
\mathbf H_q\mathbf R_b\mathbf H^H
+
\mathbf H\mathbf R_b\mathbf H_q^H,
\\
\mathbf Q_q
&=
\mathbf H_q\mathbf C_b\mathbf H^T
+
\mathbf H\mathbf C_b\mathbf H_q^T.
\end{aligned}
\]
There is no noise-covariance derivative because
$\mathbf R_n^{\rm true}$ is held fixed.

The second-order channel derivatives satisfy
\[
\mathbf H_{\tau\nu}
=
\mathbf H_{\nu\tau}
=
j2\pi\eta\mathbf D\mathbf H_\tau,
\qquad
\mathbf H_{\nu\nu}
=
-(2\pi\eta)^2\mathbf D^2\mathbf H.
\]
The entries of $\mathbf H_{\tau\tau}$ are
obtained from those of $\mathbf H_\tau$
by replacing $-T\phi_k'(\cdot)$ with
$T^2\phi_k''(\cdot)$.
Consequently,
\[
\boldsymbol\mu_{y,qr}
=
\begin{bmatrix}
\mathbf H_{qr}\boldsymbol\mu_b\\
\mathbf H_{qr}^*\boldsymbol\mu_b^*
\end{bmatrix},
\qquad
\mathbf C_{qr}
=
\begin{bmatrix}
\mathbf H_{qr}\mathbf R_b\\
\mathbf H_{qr}^*\mathbf C_b^*
\end{bmatrix}.
\]
Applying the product rule gives
\[
\begin{aligned}
\mathbf P_{qr}
&=
\mathbf H_{qr}\mathbf R_b\mathbf H^H
+
\mathbf H_q\mathbf R_b\mathbf H_r^H
\\
&\quad+
\mathbf H_r\mathbf R_b\mathbf H_q^H
+
\mathbf H\mathbf R_b\mathbf H_{qr}^H,
\\
\mathbf Q_{qr}
&=
\mathbf H_{qr}\mathbf C_b\mathbf H^T
+
\mathbf H_q\mathbf C_b\mathbf H_r^T
\\
&\quad+
\mathbf H_r\mathbf C_b\mathbf H_q^T
+
\mathbf H\mathbf C_b\mathbf H_{qr}^T.
\end{aligned}
\]
Since the differentiation variables are real,
differentiation commutes with complex conjugation.
Thus,
\[
\mathbf R_{y,q}
=
\begin{bmatrix}
\mathbf P_q&\mathbf Q_q\\
\mathbf Q_q^*&\mathbf P_q^*
\end{bmatrix},
\qquad
\mathbf R_{y,qr}
=
\begin{bmatrix}
\mathbf P_{qr}&\mathbf Q_{qr}\\
\mathbf Q_{qr}^*&\mathbf P_{qr}^*
\end{bmatrix}.
\]

For a frozen affine receiver
$\mathcal R=(\mathbf a,\mathbf W)$, let
\[
\mathbf m_{\mathcal R}
=
\boldsymbol\mu_b-\mathbf a
-\mathbf W^H\boldsymbol\mu_y.
\]
Because the coefficients are fixed,
$\mathbf m_{\mathcal R,q}
=-\mathbf W^H\boldsymbol\mu_{y,q}$.
Differentiating \eqref{eq:results_pointwise_MSE}
therefore gives
\begin{align}
M_{\mathcal R,q}
&=
-2\Re\!\left\{
\mathbf m_{\mathcal R}^H
\mathbf W^H\boldsymbol\mu_{y,q}
\right\}
\nonumber\\
&\quad
-2\Re\!\left\{\tr(\mathbf W^H\mathbf C_q)\right\}
+\tr(\mathbf W^H\mathbf R_{y,q}\mathbf W),
\label{eq:results_first_mse_derivative}
\\
M_{\mathcal R,qr}
&=
2\Re\!\left\{
\boldsymbol\mu_{y,r}^H
\mathbf W\mathbf W^H\boldsymbol\mu_{y,q}
-
\mathbf m_{\mathcal R}^H
\mathbf W^H\boldsymbol\mu_{y,qr}
\right\}
\nonumber\\
&\quad
-2\Re\!\left\{\tr(\mathbf W^H\mathbf C_{qr})\right\}
+\tr(\mathbf W^H\mathbf R_{y,qr}\mathbf W).
\label{eq:results_second_mse_derivative}
\end{align}

The gradient of $D_s(\bfeps;\boldsymbol\xi)$
is obtained by subtracting the reference-receiver
first derivatives from the plug-in WLTV
first derivatives.
The Hessian is obtained by the same subtraction
of second derivatives.
Evaluating these expressions at $\bfeps_0$
gives $\mathbf g_s$ and $\mathbf B_s$.

\section{Proof of Lemma~\ref{lem:posterior_results}}
\label{app:proof_lemma_posterior}

Recall the observation model
$\mathbf y=\mathbf H(\bfth)\mathbf b+\mathbf n$,
where
$\mathbf H(\bfth)\equiv\mathbf H(\bfeps,\mathbf h)$.
Under A1--A2,
$\E[\mathbf b\mid\bfth,\mathsf S]=\boldsymbol\mu_b$
and
$\E[\mathbf n\mid\bfth,\mathsf S]=\mathbf0$.
Hence,
\[
\E[\underline{\mathbf y}\mid\bfth,\mathsf S]
=
\begin{bmatrix}
\mathbf H(\bfth)\boldsymbol\mu_b\\
\mathbf H(\bfth)^*\boldsymbol\mu_b^*
\end{bmatrix}
=
\boldsymbol\mu_y(\bfth).
\]
The tower property gives
$\boldsymbol\mu_{\mathsf S}
=\E[\boldsymbol\mu_y(\bfth)\mid\mathsf S]$.

For the cross-covariance, decompose
\[
\underline{\mathbf y}-\boldsymbol\mu_{\mathsf S}
=
\bigl(
\underline{\mathbf y}-\boldsymbol\mu_y(\bfth)
\bigr)
+
\bigl(
\boldsymbol\mu_y(\bfth)-\boldsymbol\mu_{\mathsf S}
\bigr).
\]
Substitution into the definition of
$\mathbf C_{\mathsf S}$ yields
\begin{align*}
\mathbf C_{\mathsf S}
&=
\E\!\left[
(\underline{\mathbf y}-\boldsymbol\mu_y(\bfth))
(\mathbf b-\boldsymbol\mu_b)^H
\mid\mathsf S
\right]
\\
&\quad+
\E\!\left[
(\boldsymbol\mu_y(\bfth)-\boldsymbol\mu_{\mathsf S})
(\mathbf b-\boldsymbol\mu_b)^H
\mid\mathsf S
\right].
\end{align*}
The second term vanishes because
$\boldsymbol\mu_y(\bfth)-\boldsymbol\mu_{\mathsf S}$
is $(\bfth,\mathsf S)$-measurable and A1 gives
$\E[\mathbf b-\boldsymbol\mu_b\mid\bfth,\mathsf S]
=\mathbf0$.

For the first term, A1--A2 imply
\[
\E\!\left[
(\underline{\mathbf y}-\boldsymbol\mu_y(\bfth))
(\mathbf b-\boldsymbol\mu_b)^H
\mid\bfth,\mathsf S
\right]
=
\mathbf C(\bfth).
\]
Applying the tower property therefore gives
$\mathbf C_{\mathsf S}
=\E[\mathbf C(\bfth)\mid\mathsf S]$.

The conditional law of total covariance gives
\begin{align*}
\mathbf R_{\mathsf S}
&=
\E\!\left[
\operatorname{Cov}
(\underline{\mathbf y}\mid\bfth,\mathsf S)
\mid\mathsf S
\right]
\\
&\quad+
\operatorname{Cov}\left(
\E[\underline{\mathbf y}\mid\bfth,\mathsf S]
\mid\mathsf S
\right).
\end{align*}
Under A1--A2,
$\operatorname{Cov}
(\underline{\mathbf y}\mid\bfth,\mathsf S)
=\mathbf R_y(\bfth)$.
Together with the conditional-mean identity above,
this proves \eqref{eq:results_total_cov}.

To establish the lower bound, let
$\underline{\mathbf n}
=[\,\mathbf n^T\;\mathbf n^H\,]^T$.
Properness under A2 gives the conditional
augmented noise covariance
\[
\mathbf R_{\underline n}(\bfth)
=
\begin{bmatrix}
\mathbf R_n^{\rm true}&\mathbf0\\
\mathbf0&(\mathbf R_n^{\rm true})^*
\end{bmatrix}.
\]
By A3,
$\mathbf R_{\underline n}(\bfth)
\succeq\sigma_0^2\mathbf I_{2N}$.
The signal covariance is positive semidefinite,
and the augmented signal and noise are uncorrelated
under A2. Thus,
\[
\mathbf R_y(\bfth)
\succeq
\mathbf R_{\underline n}(\bfth)
\succeq
\sigma_0^2\mathbf I_{2N}
\qquad\text{almost surely}.
\]
Conditional expectation preserves this ordering.
Since
$\operatorname{Cov}
(\boldsymbol\mu_y(\bfth)\mid\mathsf S)
\succeq\mathbf0$,
\eqref{eq:results_total_cov} implies
\[
\mathbf R_{\mathsf S}
\succeq
\sigma_0^2\mathbf I_{2N}
\qquad\text{almost surely}.
\]
This completes the proof.

\section{Proof of Lemma~\ref{lem:affine_identity_results}}
\label{app:proof_lemma_affine}

Condition on $U$ throughout and suppress the
explicit $U$-dependence of $\mathbf a$ and
$\mathbf W$.
Define
\[
\widetilde{\mathbf b}
=
\mathbf b-\boldsymbol\mu_b,
\qquad
\widetilde{\mathbf y}
=
\underline{\mathbf y}-\boldsymbol\mu_U.
\]
Both variables have zero conditional mean, with
\[
\begin{aligned}
\E[\widetilde{\mathbf b}\widetilde{\mathbf b}^H\mid U]
&=\mathbf R_b,
\\
\E[\widetilde{\mathbf y}\widetilde{\mathbf y}^H\mid U]
&=\mathbf R_U,
\\
\E[\widetilde{\mathbf y}\widetilde{\mathbf b}^H\mid U]
&=\mathbf C_U.
\end{aligned}
\]
Let
$\mathbf d_U
=\boldsymbol\mu_b-\mathbf a-\mathbf W^H\boldsymbol\mu_U$.
The estimation error is
\[
\mathbf b-\mathbf a-\mathbf W^H\underline{\mathbf y}
=
\mathbf d_U+\widetilde{\mathbf b}
-\mathbf W^H\widetilde{\mathbf y}.
\]
The cross term between $\mathbf d_U$ and the
centered error has zero conditional expectation.
Consequently,
\[
\mathcal J(\mathbf a,\mathbf W\mid U)
=
\|\mathbf d_U\|_2^2
+
\E\!\left[
\|\widetilde{\mathbf b}
-\mathbf W^H\widetilde{\mathbf y}\|_2^2
\mid U
\right].
\]

Expanding the second term gives
\[
\E[\|\widetilde{\mathbf b}\|_2^2\mid U]
=
\tr(\mathbf R_b)
\]
and
\[
\E[\|\mathbf W^H\widetilde{\mathbf y}\|_2^2\mid U]
=
\tr(\mathbf W^H\mathbf R_U\mathbf W).
\]
Using the trace identity for a scalar product,
\[
\E[
\widetilde{\mathbf b}^H
\mathbf W^H\widetilde{\mathbf y}
\mid U]
=
\tr(\mathbf W^H\mathbf C_U).
\]
The two conjugate cross terms therefore sum to
$-2\Re\{\tr(\mathbf W^H\mathbf C_U)\}$,
proving \eqref{eq:results_affine_identity}.

For fixed $\mathbf W$, the only term depending
on $\mathbf a$ is
$\|\boldsymbol\mu_b-\mathbf a
-\mathbf W^H\boldsymbol\mu_U\|_2^2$.
Its minimum is zero, attained uniquely at
$\mathbf a=\boldsymbol\mu_b
-\mathbf W^H\boldsymbol\mu_U$.

\section{Proof of Theorem~\ref{thm:viability_results}}
\label{app:proof_viability}

Fix $s$ and $\boldsymbol\xi$.
The receivers $\mathcal R_{\rm PI}$ and
$\mathcal R_{\rm r}$ remain unchanged as
$\bfeps$ varies, so $D_s$ depends on $\bfeps$
only through the physical-state statistics.

By the assumed continuous differentiability,
Taylor expansion around $\bfeps_0$ gives
\[
\begin{aligned}
D_s(\bfeps_0+\boldsymbol\delta;\boldsymbol\xi)
&=
D_s(\bfeps_0;\boldsymbol\xi)
+
\nabla_{\bfeps}D_s(\bfeps_0;\boldsymbol\xi)^T
\boldsymbol\delta
\\
&\quad+
\frac12\boldsymbol\delta^T
\nabla_{\bfeps}^2D_s(\bfeps_0;\boldsymbol\xi)
\boldsymbol\delta
+
R_3(\boldsymbol\delta).
\end{aligned}
\]
Substituting the definitions of $\Gamma_s$,
$\mathbf g_s$, and $\mathbf B_s$ proves
\eqref{eq:results_Taylor}.
Since $\Gamma_s>0$, the expansion point
satisfies $D_s(\bfeps_0;\boldsymbol\xi)<0$
and is therefore viable.

For $\boldsymbol\delta\in\mathcal B_r$,
the segment from $\bfeps_0$ to
$\bfeps_0+\boldsymbol\delta$ lies in
$\bfeps_0+\mathcal B_r$.
The integral form of the remainder is
\[
R_3(\boldsymbol\delta)
=
\frac12\int_0^1(1-t)^2
\nabla_{\bfeps}^3
D_s(\bfeps_0+t\boldsymbol\delta;\boldsymbol\xi)
[\boldsymbol\delta,\boldsymbol\delta,\boldsymbol\delta]
\,dt.
\]
The definition of $L_3(r)$ and
$\frac12\int_0^1(1-t)^2\,dt=1/6$
therefore give
\[
|R_3(\boldsymbol\delta)|
\le
\frac{L_3(r)}{6}\|\boldsymbol\delta\|_2^3.
\]
Hence,
\[
\begin{aligned}
q_s(\boldsymbol\delta)
-\frac{L_3(r)}{6}\|\boldsymbol\delta\|_2^3
&\le
D_s(\bfeps_0+\boldsymbol\delta;\boldsymbol\xi),
\\
D_s(\bfeps_0+\boldsymbol\delta;\boldsymbol\xi)
&\le
q_s(\boldsymbol\delta)
+\frac{L_3(r)}{6}\|\boldsymbol\delta\|_2^3.
\end{aligned}
\]

For a point in $\mathcal V_{\rm in}(r)$,
the upper bound is strictly negative, so
the point belongs to $\mathcal V_s(\boldsymbol\xi)$.
Conversely, at any viable point within the ball,
the lower bound is no greater than the strictly
negative value of $D_s$.
That point therefore belongs to
$\mathcal V_{\rm out}(r)$.
These two observations prove
\eqref{eq:results_viability_sandwich}.

\section{Proof of Theorem~\ref{thm:sideinfo_results}}
\label{app:proof_sideinfo}

Apply Lemma~\ref{lem:affine_identity_results}
with $U=\mathsf S$.
For a given realization of $\mathsf S$,
the conditional MSE is
\begin{align*}
\tr(\mathbf R_b)
&+
\|\boldsymbol\mu_b-\mathbf a
-\mathbf W^H\boldsymbol\mu_{\mathsf S}\|_2^2
\\
&-
2\Re\{\tr(\mathbf W^H\mathbf C_{\mathsf S})\}
+
\tr(\mathbf W^H\mathbf R_{\mathsf S}\mathbf W).
\end{align*}
For fixed $\mathbf W$, the unique minimizing
offset is
$\mathbf a=\boldsymbol\mu_b
-\mathbf W^H\boldsymbol\mu_{\mathsf S}$.

Since $\mathbf R_{\mathsf S}\succ\mathbf0$
almost surely, the remaining quadratic expression
is strictly convex in $\mathbf W$.
Set
$\mathbf W_\star
=\mathbf R_{\mathsf S}^{-1}\mathbf C_{\mathsf S}$.
Completing the square gives
\begin{align}
&\tr(\mathbf W^H\mathbf R_{\mathsf S}\mathbf W)
-2\Re\{\tr(\mathbf W^H\mathbf C_{\mathsf S})\}
\label{eq:results_square_completion}\\
&\quad=
\tr\!\left(
(\mathbf W-\mathbf W_\star)^H
\mathbf R_{\mathsf S}
(\mathbf W-\mathbf W_\star)
\right)
\nonumber\\
&\qquad-
\tr\!\left(
\mathbf C_{\mathsf S}^H
\mathbf R_{\mathsf S}^{-1}
\mathbf C_{\mathsf S}
\right).
\nonumber
\end{align}
The first term on the right is nonnegative
and vanishes only at $\mathbf W=\mathbf W_\star$.
Together with the optimal offset, this gives
\eqref{eq:results_Wstar}.

The conditional joint covariance satisfies
\[
\begin{bmatrix}
\mathbf R_{\mathsf S}&\mathbf C_{\mathsf S}\\
\mathbf C_{\mathsf S}^H&\mathbf R_b
\end{bmatrix}
\succeq\mathbf0.
\]
Its Schur complement gives
\[
\mathbf R_b
-
\mathbf C_{\mathsf S}^H
\mathbf R_{\mathsf S}^{-1}
\mathbf C_{\mathsf S}
\succeq\mathbf0.
\]
Therefore,
\[
0
\le
\tr\!\left(
\mathbf C_{\mathsf S}^H
\mathbf R_{\mathsf S}^{-1}
\mathbf C_{\mathsf S}
\right)
\le
\tr(\mathbf R_b)
\qquad\text{almost surely}.
\]
The optimal estimate can be written as
\[
\widehat{\mathbf b}_{\mathsf S}^{\star}
=
\boldsymbol\mu_b
+
\mathbf W_\star^H
(\underline{\mathbf y}-\boldsymbol\mu_{\mathsf S}),
\]
so
\[
\E[
\|\widehat{\mathbf b}_{\mathsf S}^{\star}\|_2^2
\mid\mathsf S]
=
\|\boldsymbol\mu_b\|_2^2
+
\tr\!\left(
\mathbf C_{\mathsf S}^H
\mathbf R_{\mathsf S}^{-1}
\mathbf C_{\mathsf S}
\right).
\]
This is bounded by
$\|\boldsymbol\mu_b\|_2^2+\tr(\mathbf R_b)$,
which is finite under A4.
The minimizing estimate is therefore admissible.
Its coefficients are measurable because
conditional moments are measurable and matrix
inversion is continuous on the positive-definite
matrices.

The minimum conditional MSE is
\[
\tr(\mathbf R_b)
-
\tr\!\left(
\mathbf C_{\mathsf S}^H
\mathbf R_{\mathsf S}^{-1}
\mathbf C_{\mathsf S}
\right).
\]
Averaging over $\mathsf S$ proves
\eqref{eq:results_JSstar}.
Retaining the mean-mismatch term and the
completed-square term, subtracting the minimum
conditional MSE, and averaging proves
\eqref{eq:results_excess_risk}.
Strict positivity of $\mathbf R_{\mathsf S}$
establishes uniqueness up to almost-sure equality.

\section{Proof of Theorem~\ref{thm:oracle_results}}
\label{app:proof_oracle}

Apply Lemma~\ref{lem:affine_identity_results}
with $U=\mathsf O=(\bfth,\mathsf S)$.
Under A1--A2,
\[
\boldsymbol\mu_{\mathsf O}=\boldsymbol\mu_y(\bfth),
\qquad
\mathbf R_{\mathsf O}=\mathbf R_y(\bfth),
\qquad
\mathbf C_{\mathsf O}=\mathbf C(\bfth).
\]
Assumption A3 implies
$\mathbf R_y(\bfth)\succ\mathbf0$ almost surely.
The same square-completion argument as above gives
\[
\mathbf W_{\rm or}
=
\mathbf R_y(\bfth)^{-1}\mathbf C(\bfth),
\qquad
\mathbf a_{\rm or}
=
\boldsymbol\mu_b
-
\mathbf W_{\rm or}^H\boldsymbol\mu_y(\bfth).
\]
The corresponding Schur-complement bound ensures
a finite-second-moment estimate.
Averaging the conditional minimum MSE gives
\eqref{eq:results_Jor}.

Define the oracle residual
$\mathbf e_{\rm or}
=\mathbf b-\widehat{\mathbf b}_{\rm or}$.
The affine normal equations give
$\E[\mathbf e_{\rm or}\mid\bfth,\mathsf S]=\mathbf0$
and
\[
\E\!\left[
(\underline{\mathbf y}-\boldsymbol\mu_y(\bfth))
\mathbf e_{\rm or}^H
\mid\bfth,\mathsf S
\right]
=
\mathbf0.
\]
For fixed $(\bfth,\mathsf S)$, the difference
$\widehat{\mathbf b}_{\mathsf S}^{\star}
-\widehat{\mathbf b}_{\rm or}$
is affine in $\underline{\mathbf y}$.
The normal equations therefore imply
\[
\E\!\left[
\mathbf e_{\rm or}^H
(\widehat{\mathbf b}_{\mathsf S}^{\star}
-\widehat{\mathbf b}_{\rm or})
\right]
=
0.
\]
Expanding
\[
\mathbf b-\widehat{\mathbf b}_{\mathsf S}^{\star}
=
\mathbf e_{\rm or}
+
(\widehat{\mathbf b}_{\rm or}
-\widehat{\mathbf b}_{\mathsf S}^{\star})
\]
and using this orthogonality gives
\[
\mathcal J_{\mathsf S}^{\star}
=
\mathcal J_{\rm or}^{\star}
+
\E\!\left[
\|\widehat{\mathbf b}_{\mathsf S}^{\star}
-\widehat{\mathbf b}_{\rm or}\|_2^2
\right].
\]
This proves \eqref{eq:results_Ginfo_projection}.

To establish orthogonality between the two
estimator corrections, first decompose
\[
\underline{\mathbf y}-\boldsymbol\mu_{\mathsf S}
=
(\underline{\mathbf y}-\boldsymbol\mu_y(\bfth))
+
(\boldsymbol\mu_y(\bfth)-\boldsymbol\mu_{\mathsf S}).
\]
The first term is conditionally orthogonal to
$\mathbf e_{\rm or}$ by the oracle normal equation.
The second is $(\bfth,\mathsf S)$-measurable,
while the oracle residual has zero conditional
mean given that information.
Conditioning down to $\mathsf S$ therefore gives
\[
\E\!\left[
(\underline{\mathbf y}-\boldsymbol\mu_{\mathsf S})
\mathbf e_{\rm or}^H
\mid\mathsf S
\right]
=
\mathbf0.
\]

Likewise, the side-information-optimal residual
$\mathbf e_{\mathsf S}
=\mathbf b-\widehat{\mathbf b}_{\mathsf S}^{\star}$
satisfies
\[
\E[
(\underline{\mathbf y}-\boldsymbol\mu_{\mathsf S})
\mathbf e_{\mathsf S}^H
\mid\mathsf S]
=
\mathbf0.
\]
Since
$\widehat{\mathbf b}_{\mathsf S}^{\star}
-\widehat{\mathbf b}_{\rm or}
=\mathbf e_{\rm or}-\mathbf e_{\mathsf S}$,
subtraction gives
\begin{equation}
\E\!\left[
(\underline{\mathbf y}-\boldsymbol\mu_{\mathsf S})
(\widehat{\mathbf b}_{\mathsf S}^{\star}
-\widehat{\mathbf b}_{\rm or})^H
\mid\mathsf S
\right]
=
\mathbf0.
\label{eq:results_projection_orthogonality}
\end{equation}
Both optimal estimates have conditional mean
$\boldsymbol\mu_b$ given $\mathsf S$:
this holds directly for the $\mathsf S$-optimal
estimate and follows by the tower property for
the oracle estimate.
Their difference thus has zero conditional mean.

The plug-in correction can be written as
\[
\begin{aligned}
\widehat{\mathbf b}_{\rm PI}
-\widehat{\mathbf b}_{\mathsf S}^{\star}
&=
\mathbf d_{\rm PI}(\mathsf S)
\\
&\quad+
(\mathbf W_{\rm PI}-\mathbf W_\star)^H
(\underline{\mathbf y}-\boldsymbol\mu_{\mathsf S}),
\end{aligned}
\]
where
\[
\begin{aligned}
\mathbf d_{\rm PI}(\mathsf S)
&=
\mathbf a_{\rm PI}
+\mathbf W_{\rm PI}^H\boldsymbol\mu_{\mathsf S}
-\boldsymbol\mu_b
\\
&=
\mathbf W_{\rm PI}^H
(\boldsymbol\mu_{\mathsf S}
-\widehat{\boldsymbol\mu}_y).
\end{aligned}
\]
The constant term is orthogonal to the information
correction because that correction has zero
conditional mean.
The centered-observation term is orthogonal by
\eqref{eq:results_projection_orthogonality}.
Conditioning on $\mathsf S$ and then averaging
proves \eqref{eq:results_orthogonality}.
Here, as in the excess-risk identity, the
plug-in receiver is assumed admissible, so
these inner products are integrable.

Adding
$\mathcal G_{\rm PI}
=\mathcal J_{\rm PI}-\mathcal J_{\mathsf S}^{\star}$
and
$\mathcal G_{\rm info}
=\mathcal J_{\mathsf S}^{\star}
-\mathcal J_{\rm or}^{\star}$
proves \eqref{eq:results_total_gap}.
The established orthogonality gives this
decomposition its Pythagorean interpretation.

Finally, the explicit information-loss expression
follows by writing
\[
\begin{aligned}
\widehat{\mathbf b}_{\mathsf S}^{\star}
-\widehat{\mathbf b}_{\rm or}
&=
\mathbf W_\star^H
(\boldsymbol\mu_y(\bfth)-\boldsymbol\mu_{\mathsf S})
\\
&\quad+
(\mathbf W_\star-\mathbf W_{\rm or})^H
(\underline{\mathbf y}-\boldsymbol\mu_y(\bfth)).
\end{aligned}
\]
Conditioned on $(\bfth,\mathsf S)$, the second
term is centered and has covariance determined
by $\mathbf R_y(\bfth)$.
The cross term therefore vanishes.
Taking the squared norm and averaging gives
\eqref{eq:results_Ginfo_explicit}.

\section{Proof of Corollary~\ref{cor:zero_information_results}}
\label{app:proof_zero_info}

Equation~\eqref{eq:results_Ginfo_projection}
expresses $\mathcal G_{\rm info}$ as the
expectation of the nonnegative random variable
\[
\|\widehat{\mathbf b}_{\mathsf S}^{\star}
-\widehat{\mathbf b}_{\rm or}\|_2^2.
\]
Its expectation is zero if and only if it is
zero almost surely, which is equivalent to
equality of the two estimates almost surely.

\section{Proof of Theorem~\ref{thm:finite_state_results}}
\label{app:proof_finite_state}

Fix a state $j$ with $\Pr(Z=j)>0$.
Because $Z$ is a function of $\mathsf S$,
A1 and the tower property give
\[
\E[\mathbf b\mid Z=j]=\boldsymbol\mu_b,
\qquad
\operatorname{Cov}(\mathbf b\mid Z=j)=\mathbf R_b.
\]
The conditional affine-WL problem therefore
has the form of
Lemma~\ref{lem:affine_identity_results},
with moments
$\boldsymbol\mu_j$, $\mathbf R_j$, and
$\mathbf C_j$.
Since $\mathbf R_j\succ\mathbf0$,
square completion gives the unique coefficients
in \eqref{eq:results_finite_state_filter}.
Coefficients assigned to zero-probability
states do not affect the MSE.
The same Schur-complement argument used for
the continuous receiver establishes admissibility.

Every $Z$-dependent receiver is also an
$\mathsf S$-dependent receiver.
To obtain the exact finite-state penalty, set
\[
\mathbf d_Z(\mathsf S)
=
\mathbf a_Z^\star
+
(\mathbf W_Z^\star)^H\boldsymbol\mu_{\mathsf S}
-
\boldsymbol\mu_b.
\]
The difference between the optimal estimates is
\[
\begin{aligned}
\widehat{\mathbf b}_Z^\star
-\widehat{\mathbf b}_{\mathsf S}^{\star}
&=
\mathbf d_Z(\mathsf S)
\\
&\quad+
(\mathbf W_Z^\star-\mathbf W_\star)^H
(\underline{\mathbf y}-\boldsymbol\mu_{\mathsf S}).
\end{aligned}
\]
Conditioned on $\mathsf S$, the observation
in the second term is centered.
Hence,
\[
\begin{aligned}
&\E\!\left[
\|\widehat{\mathbf b}_Z^\star
-\widehat{\mathbf b}_{\mathsf S}^{\star}\|_2^2
\mid\mathsf S
\right]
\\
&\quad=
\|\mathbf d_Z(\mathsf S)\|_2^2
\\
&\qquad+
\tr\!\left(
(\mathbf W_Z^\star-\mathbf W_\star)^H
\mathbf R_{\mathsf S}
(\mathbf W_Z^\star-\mathbf W_\star)
\right).
\end{aligned}
\]
The right-hand side is exactly the conditional
excess MSE in
Theorem~\ref{thm:sideinfo_results}.
Averaging proves
\eqref{eq:results_finite_state_penalty}.

The class inclusion also implies
$\mathcal J_{\mathsf S}^{\star}
\le\mathcal J_{\mathcal Q}^{\star}$
for every admissible quantizer $\mathcal Q$.
Taking the infimum over
$\mathfrak Q_Q(\boldsymbol\psi)$ proves
\eqref{eq:results_feature_lower_bound}.

Finally, suppose
$\boldsymbol\psi_1=g(\boldsymbol\psi_2)$
almost surely.
Any quantizer based on $\boldsymbol\psi_1$
can be implemented using $\boldsymbol\psi_2$
by first applying $g$.
Thus, as classes of mappings from $\mathsf S$
to receiver states,
\[
\mathfrak Q_Q(\boldsymbol\psi_1)
\subseteq
\mathfrak Q_Q(\boldsymbol\psi_2).
\]
Taking the infimum over the larger class proves
\eqref{eq:results_feature_monotonicity}.

\section{Proof of Proposition~\ref{prop:lloyd_results}}
\label{app:proof_lloyd}

For a fixed codebook, assigning $\mathsf S=s$
to state $j$ incurs conditional excess MSE
$d(s;\mathbf a_j,\mathbf W_j)$.
Consequently, the overall MSE is
\[
\mathcal J_{\mathsf S}^{\star}
+
\E\!\left[
d(\mathsf S;
\mathbf a_{\mathcal Q(\mathsf S)},
\mathbf W_{\mathcal Q(\mathsf S)})
\right].
\]
The first term is independent of the assignment.
For unrestricted routing, minimizing the
distortion pointwise therefore gives
\eqref{eq:results_Lloyd_assignment}.
Because the codebook is finite, a measurable
assignment can be obtained by selecting the
smallest minimizing index whenever a tie occurs.

Conversely, for a fixed partition, the state
probabilities and events $\{Z=j\}$ are fixed.
The total MSE is the probability-weighted sum
of the conditional MSEs in the occupied cells.
Each codeword can therefore be optimized
separately.
For $\Pr(Z=j)>0$,
Theorem~\ref{thm:finite_state_results} gives
\[
\mathbf W_j^\star
=
\mathbf R_j^{-1}\mathbf C_j,
\qquad
\mathbf a_j^\star
=
\boldsymbol\mu_b
-
(\mathbf W_j^\star)^H\boldsymbol\mu_j.
\]
Zero-probability cells do not contribute to
the objective.

Each assignment step and each codeword-update
step thus minimizes the objective with the
other component fixed.
For admissible iterates with finite initial
MSE, the resulting population MSE sequence
is nonincreasing and bounded below by
$\mathcal J_{\mathsf S}^{\star}$, so its
objective values converge.
This does not imply convergence to a globally
optimal partition.

\end{document}